\documentclass[superscriptaddress,aps,preprintnumbers,amsmath,showpacs,amssymb,prd,nofootinbib,reprint]{revtex4-1}
\usepackage{bm, color} 
 \usepackage{amssymb,amsfonts,slashed,amsthm,amsmath,graphicx, soul}

\usepackage{footmisc}
\usepackage{hyperref}
\begin{document}

\renewcommand{\thesection}{\Alph{section}}

\newcommand{\mplank}{\textrm{M}_{\textrm{P}}}
\newcommand{\mg}{m_{\gamma^{\prime}}}
\newcommand{\higgs}{H_{\scriptsize \rm h}}
\newcommand{\higgst}{\tilde{H}_{\scriptsize \rm h}}
\renewcommand{\Re}{\mathrm{Re}}
\newcommand{\MF}{{\sf B}}

\newcommand{\muu}{m_{\gamma^{\prime}}}
\newcommand{\chie}{\chi_{\rm eff}}
\newcommand{\ve}[2]{\left(
\begin{array}{c}
 #1\\
#2
\end{array}
\right)
}

\renewcommand\({\left(}
\renewcommand\){\right)}
\renewcommand\[{\left[}
\renewcommand\]{\right]}

\def\ebq{\end{equation} \begin{equation}}
\renewcommand{\figurename}{Figure.}
\renewcommand{\tablename}{Table.}
\newcommand{\Slash}[1]{{\ooalign{\hfil#1\hfil\crcr\raise.167ex\hbox{/}}}}
\newcommand{\bra}[1]{ \langle {#1} | }
\newcommand{\ket}[1]{ | {#1} \rangle }
\newcommand{\beq}{\begin{equation}}  \newcommand{\eeq}{\end{equation}}
\newcommand{\bef}{\begin{figure}}  \newcommand{\eef}{\end{figure}}
\newcommand{\bec}{\begin{center}}  \newcommand{\eec}{\end{center}}
\newcommand{\non}{\nonumber}  \newcommand{\eqn}[1]{\begin{equation} {#1}\end{equation}}
\newcommand{\laq}[1]{\label{eq:#1}}  
\newcommand{\dd}[1]{{d \o d{#1}}}
\newcommand{\Eq}[1]{Eq.~(\ref{eq:#1})}
\newcommand{\Eqs}[1]{Eqs.~(\ref{eq:#1})}
\newcommand{\eq}[1]{(\ref{eq:#1})}
\newcommand{\Sec}[1]{Sec.\ref{chap:#1}}
\newcommand{\ab}[1]{\left|{#1}\right|}
\newcommand{\vev}[1]{ \left\langle {#1} \right\rangle }
\newcommand{\bs}[1]{ {\boldsymbol {#1}} }
\newcommand{\lac}[1]{\label{chap:#1}}
\newcommand{\SU}[1]{{\rm SU{#1} } }
\newcommand{\SO}[1]{{\rm SO{#1}} }

\def\({\left(}
\def\){\right)}
\def\dt{{d \o dt}}
\def\diag{\mathop{\rm diag}\nolimits}
\def\Spin{\mathop{\rm Spin}}
\def\O{\mathcal{O}}
\def\U{\mathop{\rm U}}
\def\Sp{\mathop{\rm Sp}}
\def\SL{\mathop{\rm SL}}
\def\tr{\mathop{\rm tr}}
\newcommand{\OR}{~{\rm or}~}
\newcommand{\AND}{~{\rm and}~}
\newcommand{\EV}{ {\rm \, eV} }
\newcommand{\KEV}{ {\rm \, keV} }
\newcommand{\MEV}{ {\rm \, MeV} }
\newcommand{\GEV}{ {\rm \, GeV} }
\newcommand{\TEV}{ {\rm \, TeV} }

\def\o{\over}
\def\a{\alpha}
\def\b{\beta}
\def\c{\varepsilon}
\def\d{\delta}
\def\e{\epsilon}
\def\f{\phi}
\def\g{\gamma}
\def\h{\theta}
\def\k{\kappa}
\def\l{\lambda}
\def\m{\mu}
\def\n{\nu}
\def\p{\psi}
\def\q{\partial}
\def\r{\rho}
\def\s{\sigma}
\def\t{\tau}
\def\u{\upsilon}
\def\v{\varphi}
\def\w{\omega}
\def\x{\xi}
\def\y{\eta}
\def\z{\zeta}
\def\D{\Delta}
\def\G{\Gamma}
\def\H{\Theta}
\def\L{\Lambda}
\def\F{\Phi}
\def\P{\Psi}
\def\S{\Sigma}
\def\me{\mathrm e}
\def\ol{\overline}
\def\tl{\tilde}
\def\*{\dagger}

\newcommand{\exclude}[1]{}

\def\bra{\langle}
\def\ket{\rangle}
\def\beq{\begin{equation}}
\def\eeq{\end{equation}}
\newcommand{\C}[1]{\mathcal{#1}}
\def\ov{\overline}

\title{ The Spectrum of Dark Radiation as a Probe of Reheating
}

\author{Joerg Jaeckel}
\affiliation{Institut f\"ur theoretische Physik, Universit\"at Heidelberg,
 Philosophenweg 16, 69120 Heidelberg, Germany}
\author{Wen Yin}
\affiliation{Department of Physics, Faculty of Science, The University of Tokyo, 
Bunkyo-ku, Tokyo 113-0033, Japan}

\begin{abstract}
After inflation the Universe presumably undergoes a phase of reheating which in effect starts the thermal big bang cosmology. 
However, so far we have very little direct experimental or observational evidence of this important phase of the Universe.
In this letter, we argue that measuring the spectrum of freely propagating relativistic particles, i.e. dark radiation, produced during reheating may provide us with powerful information on the reheating phase. To demonstrate this possibility we consider a situation where the dark radiation is produced in the decays of heavy, non-relativistic particles. We show that the spectrum crucially depends on whether the heavy particle once dominated the Universe or not. Characteristic features caused by the dependence on the number of the relativistic degrees of freedom may even allow to infer the temperature when the decay of the heavy particle occurred.
\noindent
\end{abstract}
\maketitle
\flushbottom

{\bf Introduction.--}
Most models of modern particle cosmology predict a reheating phase.
While there is no shortage of possible scenarios (cf., e.g.~\cite{Bassett:2005xm} for a review)
remarkably little is known for sure about this crucial event in the history of the Universe. For example, the reheating temperature could have been as low as $\sim {\rm few} ~{\rm MeV}$~\cite{Kawasaki:1999na, Kawasaki:2000en,Hannestad:2004px,Ichikawa:2006vm,DeBernardis:2008zz,deSalas:2015glj,Hufnagel:2018bjp,Hasegawa:2019jsa,Kawasaki:2020qxm,Depta:2020zbh} or it could have been much higher~$\sim 10^{16}\,$GeV (cf., e.g.~\cite{Akrami:2018odb}).
It is therefore an interesting task to garner evidence for the existence of reheating and find ways to collect information on its details.

One reason that makes it hard to probe reheating is that during this phase the Standard Model (SM) particles thermalize and therefore most information carried by them is lost.  
One way to overcome this challenge could be very weakly interacting particles that are created during reheating
and freely propagate until today. Suitable ``messengers'' may be gravitons/gravitational waves~\cite{Tashiro:2003qp,Easther:2006gt,GarciaBellido:2007af,Dufaux:2007pt, Huang:2011gf,Hebecker:2016vbl,Amin:2018kkg,Adshead:2018doq,Kitajima:2018zco,Nakayama:2018ptw,Lozanov:2019ylm,Sang:2019ndv,Adshead:2019lbr,Adshead:2019igv, Domcke:2020xmn}, axion-like particles (ALPs)~\cite{Cicoli:2012aq,Higaki:2012ar,Conlon:2013isa,Hebecker:2014gka}, right handed neutrinos~\cite{Jaeckel:2020oet} or some other very weakly interacting new particle.
Preferably, such a messenger should remain relativistic until today, thereby constituting dark radiation. The reason for this preference is that for relativistic particles it is easier to relate the energy/momentum measured in a local experiment on Earth to the one imprinted at production. In particular the energy and direction are less affected by structure formation.

In a general setup, the messengers can be produced from inflaton or modulus decays.
In a previous paper by the present authors~\cite{Jaeckel:2020oet},  the transparency condition of the Universe for the messengers, including 
neutrinos and relativistic dark particles, were clarified. 
Satisfying these conditions, they can travel over the thermal history until today, and 
the momentum distributions carry the information of the mother particles and thus can be messengers of the reheating. 

The possibility to detect ALPs originating from modulus decays was already discussed in the IAXO white paper~\cite{Armengaud:2019uso} (we also note that the authors of Ref.~\cite{Conlon:2013isa} even already calculated the energy spectrum of the ALPs and gave analytic expressions for decays taking place in either pure matter or radiation domination\footnote{Our perspective is, however, somewhat different. We want to use the flux as a probe of reheating. We therefore study the features of the flux in more detail. In particular, we will see that a relevant feature for the discrimination between a situation where the decaying particle is responsible for reheating and where it is not, is that there is a transition between a matter and a radiation dominated background evolution close to the peak of the spectrum.}). 
In~\cite{Jaeckel:2020oet} we discussed further possibilities such as right-handed neutrinos and more general messengers in dark matter, neutrino, and cosmic microwave background experiments. 

Beyond the ability to detect the messenger particles the next task is to establish their origin from reheating and to obtain additional information.
One feature of the messenger spectrum from reheating is the isotropic angular distribution which can be distinguished from particle spectra from galactic sources~\cite{Conlon:2013isa,Jaeckel:2020oet}.
However, an isotropic distribution arises as long as the decays of heavy non-relativistic particles happens much before today~\cite{Ema:2013nda, Ema:2014ufa}.
Thus, one cannot say confidently that an isotropic spectrum must be from reheating.

In this letter, we therefore ask whether we can get clearer evidence of reheating by 
carefully studying the energy spectrum of freely propagating relativistic particles, which are produced in the decays of heavy non-relativistic particles. We show that this differential flux contains information on the equation of state of the dominant energy of the Universe 
during the time when the decays occur (cf. the next section). We find the shape of this flux and point out that by precisely measuring the spectrum, we can get strong evidence for reheating.
Moreover, changes in the number of available degrees of freedom with the temperature also leave imprints in the messenger spectrum. 
Resolving these may give direct information on the reheating temperature (see further below and also Appendix~\ref{Appendix:A}). Combining this with the peak position of the spectrum~\cite{Jaeckel:2020oet} may even allow to determine the mass of the heavy particle reheating the Universe.

Before reheating the Universe may also undergo a phase of pre-heating~\cite{Traschen:1990sw,Kofman:1994rk, Shtanov:1994ce, Yoshimura:1995gc, Kasuya:1996np, Kofman:1997yn, Berges:2002cz, Mukhanov:2005sc, Dufaux:2006ee, Matsumoto:2007rd,
  Asaka:2010kv, Mukaida:2013xxa, Amin:2019qrx, Kitajima:2017peg,
  Agrawal:2018vin, Co:2018lka, Dror:2018pdh, Lozanov:2019jxc,
  Alonso-Alvarez:2019ssa, Moroi:2020bkq}. We therefore also consider the situation of the decay of a relativistic particle (cf.~Appendix~\ref{sec:preheating}).

Our analysis is independent of the specific type of messenger. The results should therefore equally apply in the case of the aforementioned or other messengers, that are 
sufficiently weakly coupled relativistic particles.\footnote{It could also apply to gravitational waves originating from particle decays~\cite{Nakayama:2018ptw} but their frequency may be too high to be detected in near future.} 
The main assumption we make is that the messenger is produced in a two-body decay from the precursor particle.
A similar analysis is also possible if the decay of the precursor is more complicated (an example is discussed in Appendix~\ref{subsec:cascade}) but we expect that the identification of features will be less clean.

In practice, of course, the detection will depend on the type of messenger realized.
While the required measurements certainly are quite challenging we can nevertheless conclude (see also Appendix~\ref{Appendix:C}) that, given the existence of suitable messengers, future measurements at observatories such as IAXO~\cite{Irastorza:2011gs,Armengaud:2014gea,Armengaud:2019uso,Abeln:2020ywv}, IceCube~\cite{2009arXiv0907.2263W,Aartsen:2014gkd,Aartsen:2014njl,Aartsen:2020aqd,Abbasi:2020jmh} or DARWIN~\cite{Aalbers:2016jon} may shed at least a little light on reheating.

\vspace{0.2cm}

{\bf Messenger flux in the expanding Universe.--} \label{sec:setup}
In this section we start by considering the simple example, where a heavy non-relativistic particle is responsible for the reheating, i.e. it was the dominant form of energy before it decays. The {primary} decay into SM particles causes the reheating, but a {rare decay will serve as the source of} our messenger. 
This situation will turn out to be clearly distinguishable from the case where the initial heavy particle only contributes sub-dominantly to the energy density of the Universe and therefore cannot be the main source of reheating.

{\bf Setup.--}
Let us consider a non-relativistic real scalar field, $\f$, with mass, $m_\f$, decaying at the cosmic time, $t\sim t_{\rm decay}$. We have in mind the case that $\f $ is a scalar modulus or inflaton, but in general, we can also consider the situation where $\f$ is a fermion. This does not change the main conclusions. 
For us a relevant feature is that the decay is two-body. This is well motivated in many models, e.g. modulus/inflaton decay into axions/ALPs~ \cite{Cicoli:2012aq,Higaki:2012ar,Conlon:2013isa,Hebecker:2014gka} or decays into right-handed neutrinos~\cite{Lazarides:1991wu,Nakayama:2011ri,King:2017nbl, Antusch:2018zvu, Guth:2018hsa, Jaeckel:2020oet}.
For our purposes this has the advantage that the produced messenger particles have a definite energy at the time of the decay instead of being distributed over phase space, thereby their energy today contains direct information on the time of decay. 

Let us assume that $\f$ has the decay channel 
\beq
\f \to \chi \chi 
\eeq
where $\chi$ is a particle that freely travels to Earth at a velocity close to the speed of light. 
The decay rate is 
\begin{equation}
  \G_{\f \to \chi\chi } = Br_{\f \to \chi\chi } \times \G_{\rm tot},  
\end{equation}
where $\G_{\rm tot}$ is the total decay width of $\f$ and $Br_{\f \to \chi\chi } \leq 1$ is the branching fraction of $\f \to \chi\chi$.

For the decay time we have,
\begin{equation}
   t_{\rm decay}=1/\G_{\rm tot}. 
\end{equation} 
We usually assume $Br_{\f \to \chi\chi } \ll 1 $ such that the {primary} decays are to SM particles\footnote{Indeed, in the case where $\phi$ is responsible for reheating we would otherwise typically have too much dark radiation~\cite{Aghanim:2018eyx,Fields:2019pfx}.}. 

The expansion rate of the Universe is given as 
\beq
\laq{Hubble}
H\equiv \frac{\dot{a}}{a}\approx \sqrt{\frac{\rho_\f+\rho_r}{3M_{\rm pl}^2}},
\eeq
where $a$ is the scale factor and $\rho_\f$ ($\rho_r$)  is the energy density of $\f$ (radiation), and 
$M_{\rm pl}\approx 2.4\times 10^{18}\GEV$ is the reduced Planck mass.
In line with our assumption $Br_{\f \to \chi\chi } \ll 1 $ we take $\rho_r$ to be dominated by SM particles and neglect the component of $\chi$. We then have, 
\begin{equation}
   \rho_r=\frac{\pi^2 g_\star }{30} T^4.
\end{equation} 
Here, $g_{\star}$ (and $g_{s\star}$ appearing soon) is the number of relativistic degrees of freedom (for entropy) in the SM for which we use the values and behavior taken from Ref.\,\cite{Husdal:2016haj}.   

{\bf Numerical results for the flux of messenger particles.--} We can obtain the time dependence of the energy density of $\f$ as well as the radiation from
\begin{align}
\laq{1}
\dot{\rho}_\f+3H\rho_\f  &= -\G_{\rm tot} \rho_\f \\
\laq{2}\dot{s}_r+3H s_r &= c[t] \G_{\rm tot} \rho_\f \,.
\end{align}
Here, $s_r=2\pi^2g_{s\star}T^3/45$ is the entropy density and
$c[t]= \frac{4\left(T g_{s\star}'+3 g_{s\star}\right)}{3 T \left(T g'_\star+4 g_\star\right)}$.  The form of $c[t]$ is obtained from $\dot{\rho}_r \simeq \G_{\rm tot} \rho_\f $ and a prompt thermalization in a much shorter period than the expansion of Universe. We have neglected the dark radiation contribution in $s_r$.
If $\G_{\rm tot} \rho_\f\to 0$, $s_r a^3$ conserves, as expected. 
From these equations we can then also obtain the time dependence of the Hubble parameter $H$. 

Using the time dependence of $\r_\f$, $\r_r$ and $H$,  we can now obtain the differential flux of $\chi$. 
We start with the momentum distribution of $\chi$. This can be  obtained from the Boltzmann equation,
\beq
\dot{f}_k \approx  H \frac{\partial f_k}{\partial \log k}+ 2\G_{\f \to \chi\chi}\delta{(k-\frac{\sqrt{m_\f^2-4m_\chi^2}}{2})} \frac{8\pi^2}{m_\phi^2}
\eeq
 where $f_k$ is the distribution function of $\chi$ of momentum $k$ (we have assumed rotational invariance). 
 This formula is justified when the occupation number of $\chi$ is not too high (c.f. Refs.\,\cite{Moroi:2020has, Moroi:2020bkq}), and $m_\chi$ is the mass of $\chi$. 
By assuming $f_k(t\ll t_{\rm decay})=0$ and neglecting the mass of $\chi$ here and hereafter, we obtain the solution 
\beq
\laq{fk}
f_k(t)\approx 32\pi^2 \frac{\G_{\f \to \chi\chi}\rho_{\f}(t')}{H(t') m_\f^4}\theta(t-t'),
\eeq
where $t'$ satisfies $a(t) k = a(t') m_\phi/2$, and $\theta$ is the step function.  
 Here, the dependence on $t'$, which in turn depends on $a$, makes explicit the connection between the spectral shape and the expansion history.

The differential flux today (at $t=t_0$) is then given by 
\beq
\laq{flux}
\frac{d^2\Phi}{d\Omega d {E}} \,=\,\ \frac{k}{E}\frac{d k}{d E} \frac{k^2}{(2\pi)^3}f_k (t_0) \simeq  \frac{E^2}{(2\pi)^3}f_E (t_0)
\eeq
where $E=\sqrt{k^2+m_\chi^2}$ is the energy of $\chi$ and we have neglected $m_\chi$ on the r.h.s. (we will also do so in the following).
Importantly one finds that the distribution function depends on the Hubble parameter $H(t')$ at the time of the decay that contributes to the flux at $k$. This will make it possible to probe the expansion history in the early Universe by precisely measuring the differential flux.

Now, we can calculate the flux by solving Eqs.\,\eq{1} and \eq{2}. Let us define CASE A and B from two initial conditions depending on whether $\f$ once dominated the Universe or not: 
\begin{align}
\nonumber
\{\rho_{\f}^{\rm ini}, \rho_r^{\rm ini}, Br_{\f\to \chi\chi}\}_{\text{A}}&=\{(10^{2}T_\f)^4, \rho_r(10^{-5}T_\f), Br\}\\
\{\rho_{\f}^{\rm ini}, \rho_r^{\rm ini}, Br_{\f\to \chi\chi}\}_{\text{B}}&
\\\nonumber&\!\!\!\!\!\!\!\!\!\!\!\!\!\!\!\!\!\!=\{10^{-10}(T_\f)^4, \rho_r(10^{2}T_\f), 2.5\times 10^{17}Br \}.
\end{align}
Here, we have fixed the branching fraction for CASE A to be $Br$. 
The shape of the spectrum does not depend on the value of $Br$ as long as it is much smaller than $1$. The overall signal strength is simply proportional to it.
In both cases we fix 
$
\Gamma_{\rm tot}= \sqrt{g_{\star}(T_\f)\pi^2T_\f^4/(90M_{\rm pl}^2)},
$
with $T_\f=200\MEV$ being the decay temperature, i.e. $\f$ decays at the same time scale in both cases. This is also roughly the reheating temperature if the decays reheat the Universe. 
We also fix $m_\f =100\TEV.$ The branching ratio in CASE B is taken to fit CASE A in the high energy region. 
The numerical result for the spectrum $Br^{-1}(d^2 \F/ d\Omega d\log_{10}E )$ is shown in Fig.\ref{fig:1}.\footnote{In our logarithmic plots we use $\frac{d^2\Phi}{ d\Omega d\log_{10}{E}}$ because this facilitates getting an impression of the total flux in an energy interval by simply multiplying the plotted value with the logarithmic width of the interval.}
Here, we divide the differential flux by $Br$. In this way our result applies to a wide range of $Br$ unless it is so high that $\chi$ becomes the dominant component of the Universe.

We find that at energies below the peak the spectrum clearly depends on whether $\f$ once dominated the Universe or not. 
In particular, the slopes of the fluxes (in {the} log-log plane) differ by $\O(1)$ before the peak. 
\begin{figure}[t!]
\begin{center}  
\includegraphics[width=75mm]{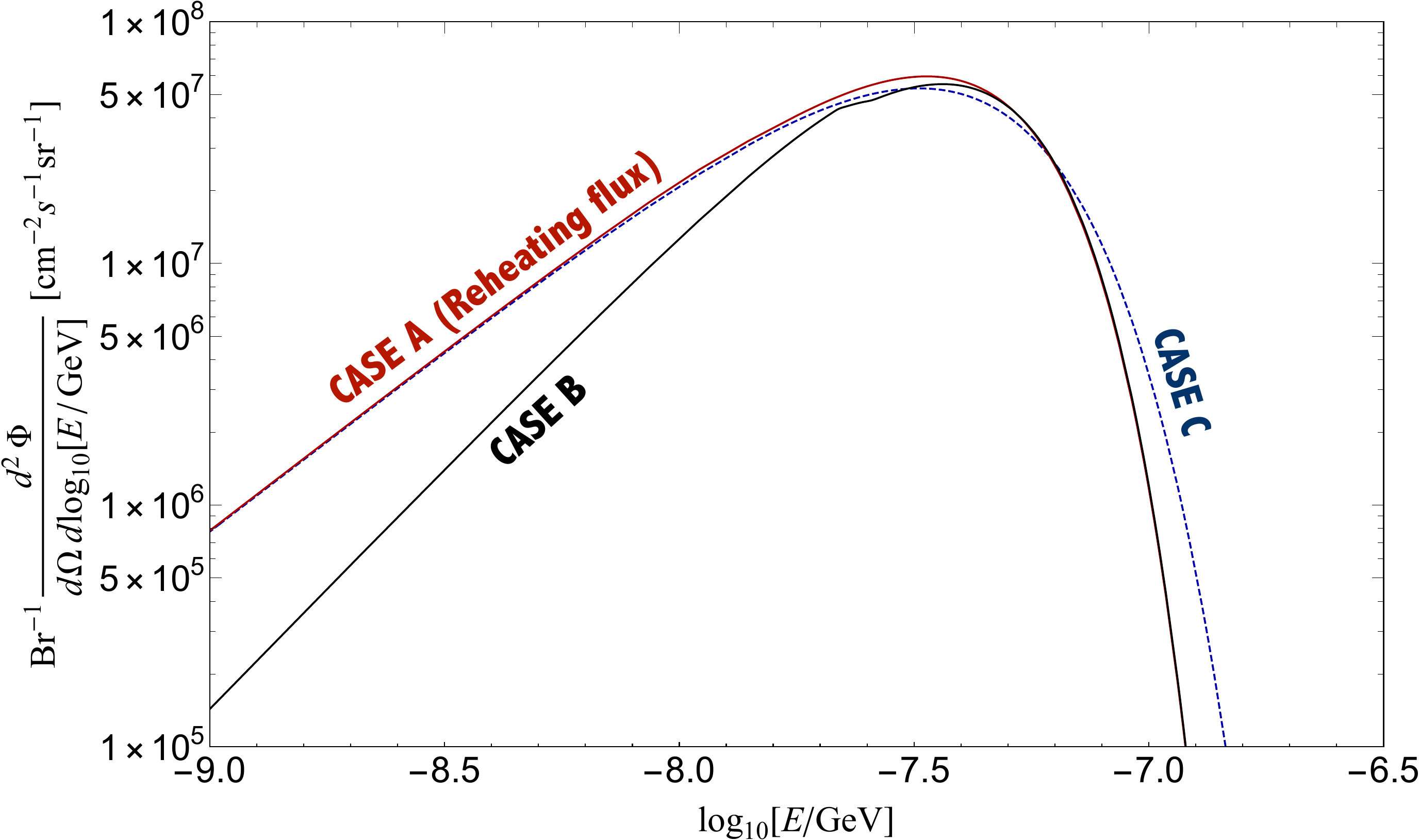}
\end{center}
\caption{The differential flux of the messenger particle, $d^2 \F/d\log_{10}{E} d\Omega$. CASE A ($\phi$ once dominated the Universe)  and CASE B ($\phi$ never dominates the Universe and decay in the radiation dominant epoch) are shown in red and black lines, respectively.
We also show the flux for CASE C where a subdominant $\f$ decays in the matter dominant era as the blue dashed line. }
\label{fig:1}
\end{figure}

As we will see from the analytical estimates below, the slope of the flux below the peak depends mainly on the equation of state of the dominant energy form. Nevertheless, the flux in CASE A can be also distinguished from the case of a subdominant $\f$ decaying during a matter dominant epoch.
This is shown as the blue dashed line (CASE C) in Fig.\,\ref{fig:1} where the decay rate of $\f$ satisfies $\G_{\rm tot}= H|_{z=100}$ i.e. in the matter dominant era. 
We use  $\frac{d^2\F}{d\Omega dE }=Br C_{\rm mat} (E H(t[z(E)]))^{-1} e^{-\Gamma_{\rm tot} t[{z(E)}]},\,\,z(E)=m_\phi/2E-1$ and the forms of $H$ and $t[z]$ given in Ref.~\cite{Jaeckel:2020oet}. 
(This form can be obtained by solving \Eq{1} and then using \Eqs{fk} and \eq{flux} while taking $H$ in the matter dominated era and neglecting the energy density of $\f$.)
We take $C_{\rm  mat}\approx 6.1\times 10^{-42}\GEV^4, m_{ \f}=5\KEV.$
to match the IR flux to the one of CASE A.
We can see that with $E$ around or above the peak energy, we have more than $\O(10)\%$ flux differences. 
More drastic differences may be observed if $\f$ decays in the dark energy dominant epoch. 
In particular, if $\f$ is decaying today (and has been non-relativistic for a sufficiently long time),  $\f$ tends to gather around the galactic center due to the gravitational interaction. 
In contrast to CASE A this flux has an angular dependence that (amongst other things) superimposes a narrow peak at $m_{\phi}/2$.

{\bf Analytical approach to the flux.--} 
To better understand the different behaviors and the ways to distinguish the different scenarios let us consider some analytical estimates.
Note that $f_E$ for a given energy $E$ is proportional to $H^{-1}$ at the scale factor $a=2E/m_\phi$. 
$H$ depends on the dominant energy of the Universe at $a$, which means that the differential fluxes, or spectral intensities,  at different $E$ scan the dominant energy of the Universe at different $a$.

Let us consider the epoch before the typical decay time $t<t_{\rm decay}= 1/\Gamma_{\rm tot}$, i.e. $a\ll a_{\rm decay}$ with $a_{\rm decay}$ being the scale factor at $\G_{\rm tot}=H$.  
Since $f_E \propto a^{-3+3/2(1+w)}$ in this epoch we obtain (in agreement with~\cite{Conlon:2013isa})
$\frac{d^2\F}{ d\Omega d E} \propto E^{\frac{1}{2}+\frac{3}{2}w}$,
where $w$ is the equation of state of the Universe, e.g. $w=0$ (1/3) for matter (radiation) dominated Universe. 
Therefore,  the slopes of the fluxes (in the log-log plane) differ by $\O(1)$ for CASE A and B when $E<E_{\rm decay}=m_\f/(2a_{\rm decay})$. 
$\chi$ with $E>E_{\rm decay}$, are produced from the few remaining $\f$ after the typical decay time. Accordingly, the flux gets an additional exponential suppression.

If $\f$ reheats the Universe, i.e. it dominates the Universe before the decay and transfers most of its energy into radiation around and after the decay, 
the differential flux of $\chi$ has two typical regimes 
\begin{align}
\frac{d^2\F_{\rm reheating}}{d \Omega d E} 
\propto 
\left\{\begin{array}{cc} 
 E^{\frac{1}{2}} ~~~~~~~~~~~~~~~~&(E\ll E_{\rm decay})\\  E e^{-\kappa(E/E_{\rm decay})^2 }& (E \gg E_{\rm decay})
\end{array}\right.\laq{rehflux}
\end{align}
where $\kappa$ is an order 1 numerical coefficient.

Such a reheating flux can also be distinguished from the decay of a subdominant species during a matter dominated phase of the Universe (CASE C). The reason is that if the decaying particle is responsible for reheating, decays after the typical decay time occur during a radiation dominated epoch. For a subdominant species such a changeover at the decay time usually does not happen.\footnote{Such a changeover at the same time as the decay would be an unlikely coincidence.} This is then reflected in the spectrum at energies higher than the peak energy, where we have (again in accord with~\cite{Conlon:2013isa})
$ \frac{d^2\F_{\rm rad}}{ d\Omega d E } \propto  E e^{-\kappa'(E/E_{\rm decay})^2 }$  
in the radiation dominant epoch and  
 $ \frac{d^2\F_{\rm mat}}{ d\Omega d E } 
\propto  E^{1/2} e^{-\kappa'' (E/E_{\rm decay})^{3/2} }$
in the matter dominant epoch. 
Again, $\kappa' \AND \kappa''$ are $\O(1)$ numerical coefficients. 

Note that the relative shapes of the differential flux do not depend on parameters of the model such as {$Br_{\rm \phi \to \chi\chi}$} or $m_\f$. Thus the reheating flux shape is a robust prediction of reheating caused by non-relativistic particle decays to relativistic messengers. If we can identify this flux, it should be strong evidence of reheating. 

{\bf Measurement of reheating parameters.--}
Let us now outline a strategy for an ultimate possibility to measure reheating parameters in the future.  
A slight change of the equation of state of the Universe will also be induced by the decoupling of the relativistic 
components in the radiation dominant era. Since soon after the reheating the Universe is radiation dominated, this effect affects the reheating flux shape. 
We depict a comparison of reheating fluxes in 
Fig.\,\ref{fig:flux_compare} with and without taking account of this decoupling effects in red-solid and gray-dotted lines, respectively, with 
$T_\phi=400\MEV$. The CASE A ($T_{\phi}=200\,{\rm MeV}$) flux is also shown in the blue-dashed line. The decoupling affects the flux via the changes of $g_{\star},g_{s \star}$, which is especially significant around the QCD phase transition. (See more details in Appendix~\ref{Appendix:A})
Again these flux shapes do not depend on  $Br$ and $m_\phi$.  Neglecting the decoupling effect (gray-dashed line) the shape even does not depend on $T_\phi.$

\begin{figure}[t!]
\begin{center}  
\includegraphics[width=75mm]{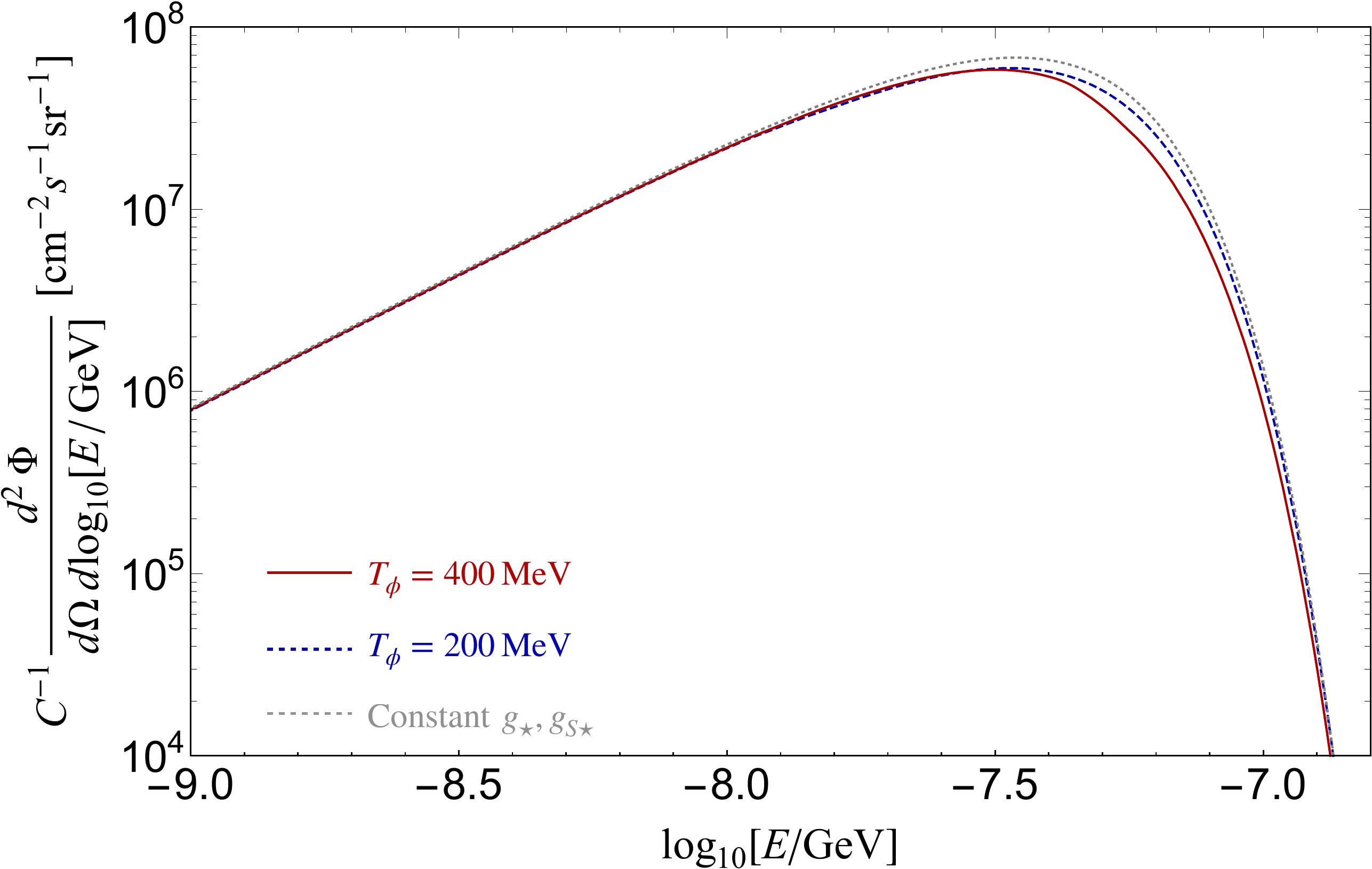}
\end{center}
\caption{The reheating flux dependence on the decoupling effect:  $T_\phi=400\MEV$ (red-solid line) and $T_\phi=200\MEV$ (blue-dashed line, CASE A). We take $g_\star,g_{s \star}$ temperature independent for the gray-dotted line. (See Appendix~\ref{Appendix:A} for details.)  }
\label{fig:flux_compare}
\end{figure} 
Conversely by carefully measuring the spectral shape, 
the reheating temperature $\approx T_\phi$ can be obtained in {principle. For example, in Fig.~\ref{fig:flux_compare} we can infer that the flux indicated by the red solid line originates from around QCD phase transition era. 
In addition, we can then obtain $m_\f$ via a determination of $m_\f/T_{\phi}$ from the peak position~\cite{Jaeckel:2020oet}. Then, $T_\phi$ can be translated into $\G_{\rm tot}$ as well as $\rho_r$ and thus $\rho_\f$ from energy conservation. From the intensity of messenger flux we can also derive $Br_{\phi \to \chi\chi}$.

\vspace{0.2cm}
{\bf Conclusions and discussion.--}\label{sec:conclusions}
Reheating is a central part of our current modelling of the early Universe.
However, it has not yet been confirmed by  direct experiment/observation. 

In this letter, we have studied the spectra of freely propagating relativistic particles produced by the decays of heavy non-relativistic particles and shown that the spectra depend significantly on whether the heavy particle once dominated the Universe. 
We have demonstrated that the energy spectrum of relativistic messengers from the decay of a non-relativistic particle exhibits clear features that can tell us whether the decaying particle was responsible for reheating. These imprints arise via the equation of state. If the spectrum can be measured with sufficient precision we may even be able to tell the reheating temperature and the mass of the decaying particle. 
One may also wonder what happens if the mother particle is relativistic, featuring its own non-trivial spectrum. In this case (cf. Appendix~\ref{sec:preheating} where we consider typical spectra arising in preheating scenarios) one cannot easily tell whether the decaying particle dominated the energy density or not. However, 
it is nevertheless usually distinguishable from the case of a non-relativistic mother particle. Furthermore, even if the messengers arise from a cascade of two subsequent two body decays, some small traces of the phase during which they originated may still be visible (cf. Appendix~\ref{subsec:cascade}). 

Our discussion here can apply to various reheating scenarios with mother particles coupled to light-weakly coupled messenger particles, such as gravitons, ALPs, dark matter, neutrinos, etc..

Let us conclude with some comments on the experimental opportunities. 
The discrimination between CASE A and B is relatively straightforward, since the flux at energies below the peak are quite different in a large range.
In Appendix~\ref{Appendix:C}, we argue that, even if we can just measure the flux in two bins with a relative width of $25\%$ corresponding to a quite moderate energy resolution, the discrimination is possible at the $2\s$ level with $\O(1000)$ events, possibly even significantly less if the data is used more efficiently than in our simplistic estimate.
This gives us an optimistic expectation for suitable experiments such as IceCube, IAXO and DARWIN~\cite{Irastorza:2011gs,Armengaud:2014gea,Armengaud:2019uso,Abeln:2020ywv,2009arXiv0907.2263W,Aartsen:2014gkd,Aartsen:2014njl,Aartsen:2020aqd,Aalbers:2016jon} (see also \cite{Arguelles:2019xgp}).

The discrimination between a reheating flux and sub-dominant $\f$ decays in a matter dominated epoch as well as a measurement of the reheating temperature is more difficult due to the exponential suppression of the flux at high energy. A good energy resolution (expected for at least some of the above mentioned experiments) will be critical to do this.
As an additional check of the possible reheating origin, the angular distribution of the flux can also play a useful role~\cite{Conlon:2013isa,Ema:2013nda,Ema:2014ufa,Jaeckel:2020oet}.
The last question is whether enough events can be observed. 
In fact, 
IceCube~\cite{Aartsen:2017mau, Aartsen:2018vtx,Abbasi:2020jmh}, ANITA~\cite{Gorham:2016zah, Gorham:2018ydl} and XENON1T~\cite{Aprile:2020tmw}
experiments have already observed anomalous events that may be from BSM physics.\footnote{ See Refs.\,\cite{Feldstein:2013kka,Esmaili:2013gha,Ema:2013nda, Higaki:2014dwa,Rott:2014kfa, Ema:2014ufa,Dudas:2014bca,Murase:2015gea,Dev:2016qbd,Hiroshima:2017hmy,Bhattacharya:2014yha,Kopp:2015bfa,Cui:2017ytb, Cherry:2018rxj, Yin:2018yjn, Fox:2018syq, Heurtier:2019git, Kannike:2020agf,Fornal:2020npv,Su:2020zny,Bloch:2020uzh, Anchordoqui:2021dls} for BSM explanations with relativistic particles.} 
As an example we show in Fig.~\ref{fig:ICECUBE}  (in Appendix~\ref{subsec:cascade}) fluxes for the different cases discussed in this letter fitting the events~\cite{Abbasi:2020jmh} of IceCube. Aside from inviting intriguing speculation this also suggests that a sufficiently good measurement to distinguish the fluxes may be feasible in the not too distant future.
If these hints persist and are not explained by other effects, we may be in the fortuitous situation that measuring their energy dependence may allow us to get a glimpse of the beginning of the thermal history.

\section*{Acknowledgments}
WY was supported by JSPS KAKENHI Grant Number 16H06490 and 19H05810.

\appendix

\section{Dependence of the flux on the number of degrees of freedom (D.O.F) }
\label{Appendix:A}

The shape of the flux also depends on the changes in the number of relativistic degrees of freedom. This becomes noticeable for $\f$ decays that happen in or close to the radiation dominated era. This holds because when radiation is a relevant part of the total energy density
changes in the number of D.O.F. become noticeable in the Hubble expansion. 
Indeed, the flux is proportional to $1/(s H)\propto g_\star^{-1/2}g_{s \star}^{-1}$ from Eq.~\eq{fk}. 
The effect can be clearly seen as the slight kinks near the peak of the spectrum of CASE B in Fig.~\ref{fig:1}. The changing of the slope just before the peak reflects the changes of $g_\star,g_{s\star}$ due to the QCD phase transition.

This decoupling effect is also important for the reheating flux as discussed in the main part since soon after reheating radiation is dominant. 
The time evolution of $T$ and $\rho_\phi$ in CASE A is shown in Fig.\ref{fig:0}.
\begin{figure}[t!]
\begin{center}  
\includegraphics[width=75mm]{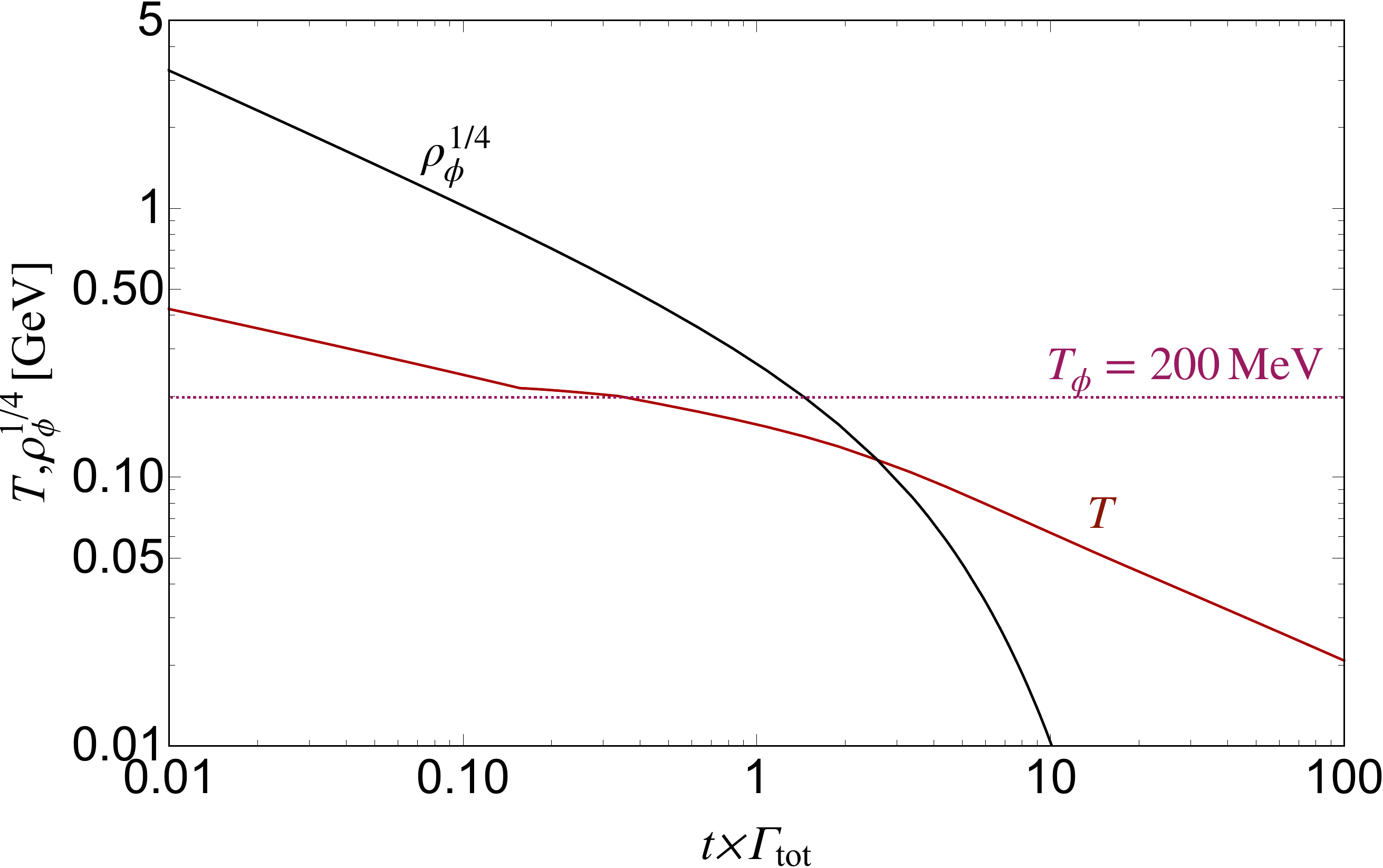}
\end{center}
\caption{Time evolution of $T$ (red line) and $\rho_\phi$ (black line) for CASE A.}
\label{fig:0}
\end{figure}
{When $t\times  \G_{\rm tot}\sim 1$ reheating ends which also roughly corresponds
to the time when the peak of the reheating flux is created.}
We see that at $T\sim 200\MEV$ the QCD phase transition slightly slows down the decrease of $T$. 
This is different from the flux with 
constant $g_\star \AND g_{s\star}.$ 
A comparison of the fluxes is shown in Fig.~\ref{fig:flux_compare}.
Here the flux divided by a parameter $C$ is shown. The CASE A flux is given in blue dashed line ($C=Br$) and the case with constant $g_\star  \AND g_{s \star}$ is indicated by the gray dotted line. For the gray dotted line, $T_\phi=400\,$MeV, $C=1.5 Br$ and $m_\phi=240\,$TeV  as well as assuming constant $g_\star(T)=g_\star(T_\phi)\approx 66.9$ and $g_{s\star}(T)=g_{s\star}(T_\phi)\approx 66.3$. We have also checked that the shape does not change when changing $T_\phi, Br, m_\f$ as long as we take constant $g_\star \AND g_{s\star}$, as can be expected. However, the peak energy and flux intensity may change. 
The difference can be more significant if $T_\f$ is slightly larger so that the QCD phase transition happens mostly in the radiation dominant epoch but with not too suppressed $\rho_\f$. The red solid line represents
$C= 1.21 { Br}$ with $T_\phi=400\MEV, m_\phi=221\TEV$. 
As we can see the fluxes can differ noticeably.
By carefully measuring the energy scale and the size of the depression, in principle, we can identify $T_\phi$.

This  discussion applies not only to the reheating flux but also in the more general case where we can measure $T_\phi$ given a flux from $\f$ decays in the radiation dominated epoch.  By carefully measuring the flux at two energies, we can in principle measure  $g^{-1/2}_{\star}g^{-1}_{S\star}$ at two different temperatures around $T_\phi$. Then $T_{\f}$ can be obtained under the assumption that the SM accounts for the D.O.F.

\section{Messenger flux from the decay of relativistic particles}\label{sec:preheating}

\subsection{Setup - Relativistic particles from preheating}
If the reheating is caused by the decay of an oscillating scalar field with a non-parabolic potential, $\f$ may be ``preheated''~\cite{Traschen:1990sw,Kofman:1994rk,Shtanov:1994ce, Yoshimura:1995gc, Kasuya:1996np,Kofman:1997yn,Berges:2002cz} and have a non-trivial spectrum before the decay. 
Let us consider the $\f$ potential
\beq 
V=\frac{m_\f^2}{2} \f^2 +\frac{\lambda}{4} \f^4
\eeq 
and assume a large enough initial field (or tiny enough mass term). 
When the amplitude is large, parametric resonance~\cite{Traschen:1990sw,Kofman:1994rk,Shtanov:1994ce, Yoshimura:1995gc, Kasuya:1996np,
  Kofman:1997yn,Berges:2002cz}\footnote{See Refs.~\cite{Mukhanov:2005sc, Dufaux:2006ee, Matsumoto:2007rd,
  Asaka:2010kv, Mukaida:2013xxa, Amin:2019qrx, Kitajima:2017peg,
  Agrawal:2018vin, Co:2018lka, Dror:2018pdh, Lozanov:2019jxc,
  Alonso-Alvarez:2019ssa,Moroi:2020bkq} for some recent studies.} becomes important. 
Soon afterwards the system enters into a turbulence regime in which a scalar field spectrum follows a power-law~\cite{Khlebnikov:1996mc,Micha:2002ey,Micha:2004bv,Lozanov:2017hjm}. 
For a sufficiently long period, almost no homogeneous mode of $\f$ remains. The relativistic $\f$ particle spectrum undergoes a self-similar evolution until the decays. A typical expectation for the occupation number is~\cite{Khlebnikov:1996mc,Micha:2002ey,Micha:2004bv,Lozanov:2017hjm} $f_{\f,k}\propto k^{-3/2}$. More generally we can consider a power-law spectrum of the form, 
\begin{equation}
f_{\f,k}\propto k^{-n}.
\end{equation}
We will now focus on the phase when such a power-law has been established, the relativistic $\f$ dominates the energy density and its decay into SM particles reheats the Universe.

Via couplings to the SM particles and $\chi$, $\f$ can decay both into SM particles and $\chi$s. 
Again we assume a two body decay to $\chi$,
\beq \f \to \chi\chi.\eeq 
Moreover, we assume that these couplings are so small that the decay processes can be treated in perturbation theory.  
The decay rate of the $k$ mode of $\f$ is 
\beq
\Gamma_{\rm tot}[k]= \frac{m_\f}{\sqrt{k^2+m_\f^2}} \times \G_{\rm tot},
\eeq
where the pre-factor is the Lorentz factor, and again $\G_{\rm tot}$ is the total decay width in the rest frame. 
We can get the decay rate to a $\chi$ pair as
\beq
\Gamma_{\f\to \chi\chi}[k]=  Br_{\f \to \chi\chi} \times \Gamma_{\rm tot}[k].
\eeq
Assuming $k\gg m_\f$, we {have} $\Gamma_{\f\to \chi\chi}[k],\Gamma_{\rm tot}[k]\propto k^{-1}$.

\subsection{Numerical result}
The equations for the time dependence of $\rho_\f$ and $\rho_r$ (neglecting the small $\chi$ density), are
\begin{align}
\laq{boltz1}
 \dot{\rho}_{\f, k}-Hk \frac{\partial {\rho_{\f, k} }}{\partial k}+4H \rho_{\f, k} &=- \Gamma_{\rm tot}[k] \rho_{\f, k} \\
\laq{boltz2}
\dot{s}_r+3H s_r &= c[t] \int_{-\infty}^{\infty}{d \log k\G_{\rm tot}[k] \rho_{\f, k}}[t]
\end{align}
where 
$
\rho_{\f,k}[t]\equiv  \frac{k^3 \sqrt{k^2+m_\f^2}}{2\pi^2}f_{\f, k} [t]
$
satisfying 
$
\rho_\f[t]=\int_{-\infty}^{\infty}{d\log 
k}\rho_{\f,k}[t].
$
$H$ is given in \eq{Hubble}. The relevant modes of $\f$ are assumed to behave as radiation. Indeed, hereafter we mostly neglect $m_\f,m_\chi$.

The $\chi$ differential flux can be obtained by assuming a subdominant decay to $\f\to \chi\chi$ with $Br_{\f \to \chi\chi}\ll 1$. 
In addition to \Eqs{boltz1} and \eq{boltz2}, we  can solve 
\begin{align}
\non
 \dot{\rho}_{\chi,k}-& Hk \frac{\partial {\rho_{\chi,k}}}{\partial k}+4H \rho_{\chi,k}\\
 &=Br_{\f \to \chi\chi}\int_{-\infty}^{\infty}{d \log k' P[k, k'] \Gamma_{\rm tot}[k'] \rho_{\f, k'}},\laq{chiboltz}
\end{align}
where $\rho_{\chi,k}$ is defined by $\rho_{\chi,k} \equiv  \frac{k^3 \sqrt{k^2+m_\chi^2}}{2\pi^2}f_{\chi,k}[t]$
and 
$
P[k, k']= 2 (k/k')^{2}\theta{(k'-k)}
$
represents the  phase space distribution of $\chi$ from a relativistic $\f$ decay with $m_\phi\gg m_\chi$. 
We get the differential flux, 
\beq
\frac{d^2\Phi}{ d\Omega d E }\,=\, \frac{\rho_{\chi, k}(t_0)}{4\pi E k}.
\eeq
These equations can be applied to the $\chi$ flux for any initial relativistic spectrum of $\f$ decaying into SM radiation and $\chi$.

Now let us come back to the preheating scenario discussed above. The initial conditions for the reheating from the self-resonant $\f$ decays are 
\beq
\rho_{\phi,k}(t_{\rm ini})= \theta{[k-k_{\rm min}]} C_X T_{\f}^4 {(k/k_{\rm max})}^{-n+4}e^{- k/k_{*}},
\eeq
with $\rho_r(t_{\rm ini}),\rho_{\chi,k}(t_{\rm ini})\sim 0$
where $C_X\gg1$ so that before the decay $\f$ particles dominate the Universe. The exponential in $\rho_{\phi,k}(t_{\rm ini})$ represents the decay of the power-law set by hand,  
and $k_{\rm min, max}$ is set for the convenience  of calculation. 
We also set 
\beq
\Gamma_{\rm tot}[k_{*}]\equiv \sqrt{g_{\star}(T_\phi)\pi^2T_\phi^4/(90M_{\rm pl}^2)},
\eeq
by which we define the reheating temperature. 
For convenience we use 
\beq \Gamma_{\rm tot}[k]\equiv (k_*/k)\Gamma_{\rm tot}[k_*].\eeq
In Fig.\,\ref{fig:reh}, we show the resulting cosmological temperature (red solid line) and the energy density of $\rho_\f$ (black solid line). 
\begin{figure}[t!]
\begin{center}  
\includegraphics[width=75mm]{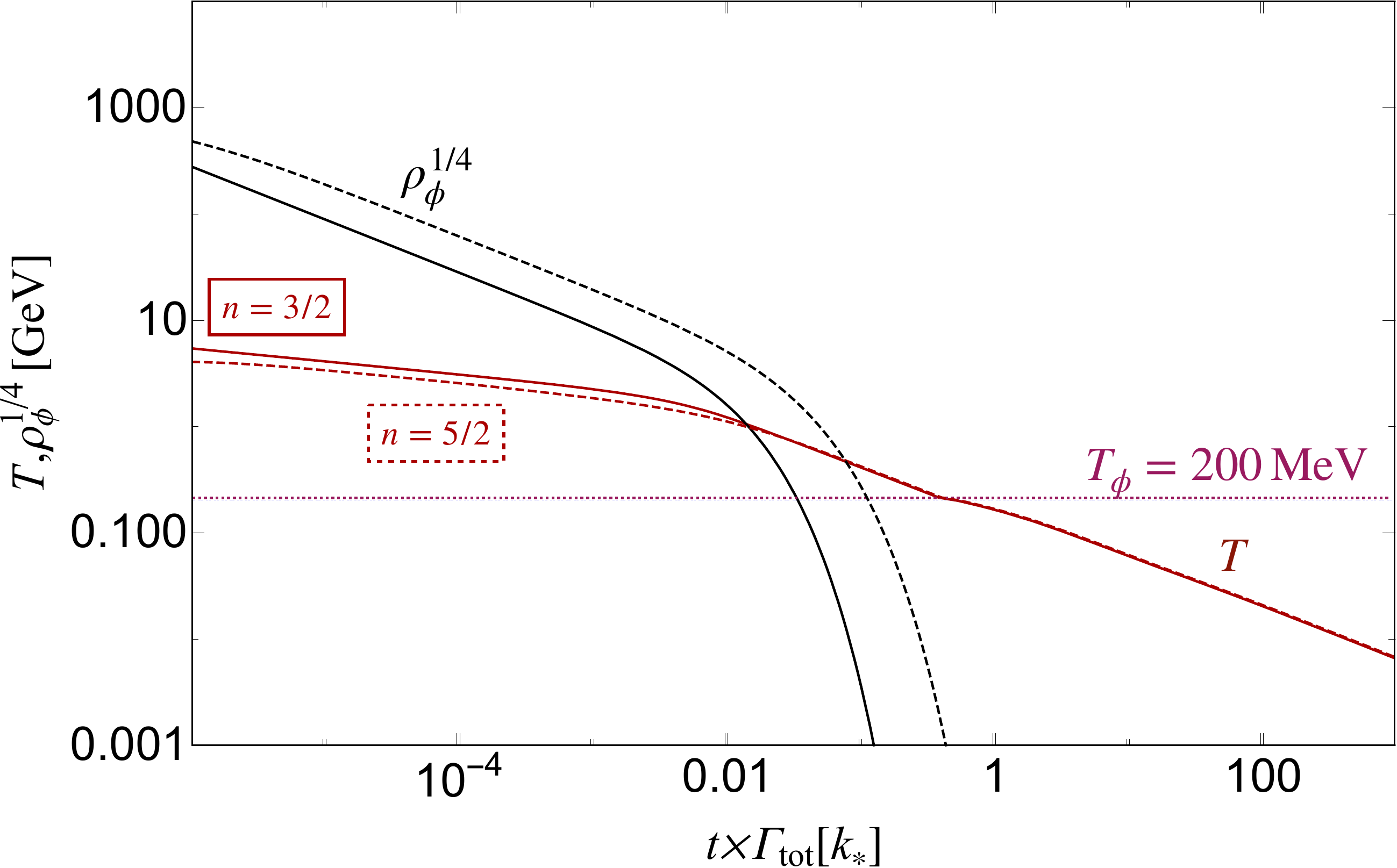}
\end{center}
\caption{$T$ (red) and $\rho_\f$ (black) as a function of $t$ in the relativistic $\f$ scenario with the occupation numbers $f_{\f,k}\propto k^{-n}$ for $n=3/2$ (solid lines) and $n=5/2$ (dashed lines). In both cases we fix $C_X=10^8, T_\phi=200\MEV, k_{*}=10^{10}\MEV, k_{\rm max}=10^{12}\GEV, k_{\rm min}=10^{7}\GEV,$ and $t_{\rm ini} \Gamma_{\rm tot}[t_*]=10^{-8}$. We neglect the masses of $\f$ and $\chi$ compared with $k$.
}
\label{fig:reh}
\end{figure}

The corresponding differential flux,
$\frac{d^2 \Phi}{d\Omega d\log_{10}{E}} \times Br_{\f \to \chi\chi}^{-1}$, is plotted in Fig.~\ref{fig:flux2}.
For comparison, we also show the initial flux of $\f$. One can see that the $\chi$ flux mimics the original $\f$ flux for $n=5/2$ but not for $n=3/2.$

\begin{figure}[t!]
\begin{center}  
\includegraphics[width=75mm]{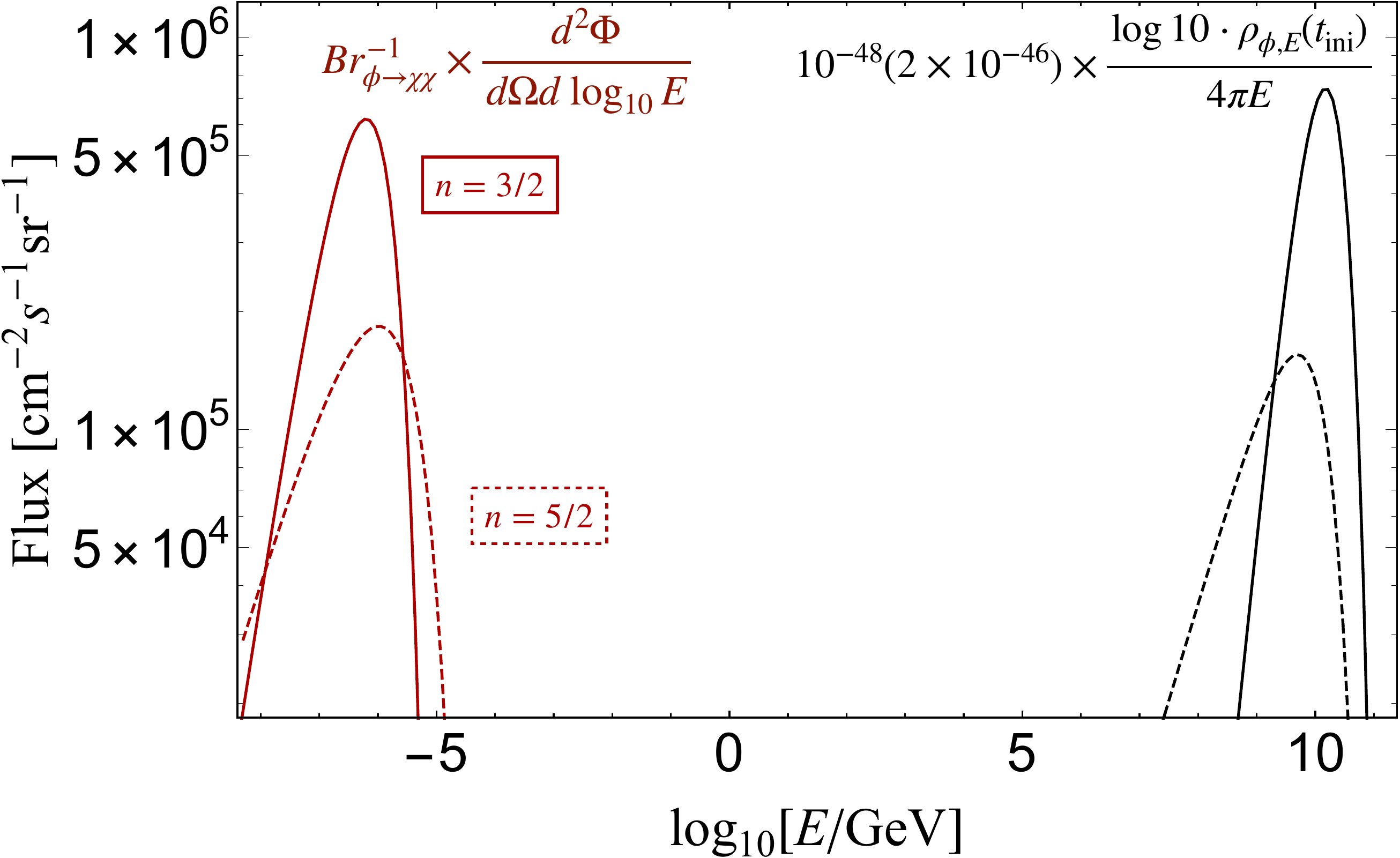}
\end{center}
\caption{The $\chi$ flux  today from relativistic $\f$ decays (red curve). 
The black curve denotes the original differential $\f$ flux, at $t=t_{\rm ini}$.
The solid (dashed) lines represent the case of $n=3/2$ ($5/2$).
The parameters are the same as those in Fig.\,\ref{fig:reh}.
}
\label{fig:flux2}
\end{figure}
\subsection{Analytical understanding and general features of the flux from relativistic particle decays}

The behavior of the $\chi$ flux for different $n$
can be understood from the equation 
\begin{align}
 \dot{\hat{\rho}}_{\chi,\hat{k}}&=Br_{\f \to \chi\chi} \int{d \log \hat{k}' P[\hat{k}, \hat{k'}]\Gamma_{\rm tot}[k'] \hat{\rho}_{\f,\hat{ k}'}} \\
 &= -Br_{\f \to \chi\chi} \int{d \log \hat{k}' P[\hat{k}, \hat{k'}]\dot{ \hat{\rho}}_{\f,\hat{ k}'}},
\end{align}
which is derived from \Eq{chiboltz}, by defining $\hat{\rho}_{\chi,\hat{k}}\equiv a^4{\rho}_{\f, k}$ and  $\hat{\rho}_{\f,\hat{k}}(t)\equiv a^4{\rho}_{\f, k}$  with $\hat{k}\equiv a \times k$ being the comoving momentum. In the second equality we have used 
$\hat{\rho}_{\f,\hat{k}}(t)=\exp{\left[-\int_{t_{\rm ini}}^t{dt'\Gamma_{\rm tot}(k') }\right]}\hat{\rho}_{\f, \hat{k}}(t_{\rm ini}).$ 
Integrating over time we have 
\beq
\laq{Boltzcom}
\hat{\rho}_{\chi,\hat{k} }(t)\simeq Br_{\f \to \chi\chi} \int{d \log \hat{k}' P[\hat{k}, \hat{k'}] \hat{\rho}_{\f,\hat{ k}'}(t_{\rm ini})},
\eeq
where we have assumed $t$ much larger than the typical decay time.
Due to the combination of the step-function in $P[\hat{k},\hat{ k}']$ and the power of $\hat{k'}$ in the integrand, 
the slope of $\rho_{\chi,\hat{k}}$ at the IR end can be approximated as 
\begin{align} 
\label{eq:slopes}
\hat{\rho}_{\chi,\hat{k} }(t)/Br_{\f \to \chi\chi}
\propto  \left\{
\begin{array}{cc} 
    \hat{\rho}_{\f,\hat{ k}}(t_{\rm ini})\propto   \hat{k}^{-n+4} &(n> 2) \\  \hat{k}^2  ~~~~~~~~~~~~~~~& (n \leq 2)
\end{array}\right..
\end{align}
Here, we have used that 
the $\hat{k'}$ integral is dominated by values of $\hat{k'}$ around $\hat{k}$ for $n>2$. In contrast for  
$n\leq 2$
the integral is dominated by the UV cutoff (for $n=2$ only logarithmically) 
and thus
approximately independent of $k$. 
Therefore, in this case $\hat{\rho}_{\chi,\hat{k}}$ is proportional to
$\hat{k}^2$ from $P[\hat{k},\hat{k}']$.
This agrees well with the numerical results.

Translating this into the flux, 
we obtain a maximal slope, $d^2\F/d\Omega d E\propto \rho_{\chi,k}/k^2\propto E^0$. As a result, the reheating flux ($\propto E^{1/2}$) cannot be mimicked by the  decays of relativistic particles with a simple power-law. 

From \Eq{Boltzcom} we can also see that the form of $\rho_{\chi,k}$ does not depend on $H$ in the early Universe as long as all $\f$ particles decay. This is  different from the case of non-relativistic $\f$ decays. 
We therefore can neither distinguish if $\f$ dominated the Universe nor infer the behavior of $H$ by the messenger flux from relativistic $\f$ decays.

\subsection{An example with a cascade decay}\label{subsec:cascade}
Another plausible scenario for the origin of a relativistic mother particle of the messenger is a cascade decay.
Let us consider a situation where a non-relativistic particle undergoes a 
two-body decay which is followed by another two-body decay to our messenger particle.
The spectrum of the relativistic intermediary particles is that of the messengers considered in the main text. From this we find that a decay in the matter (radiation) dominated phase corresponds to $n=3/2$ ($n=1$) {for energies far below the peak}. From Eq.~\eqref{eq:slopes} we can now see that the resulting spectrum for the final messengers is then independent of the original slope $n$. 
The second decay, unfortunately obscures the information on the slope. 

However, there are still some subtle traces of the expansion history/slope left. The reason is that the $k^2$ behavior of Eq.~\eqref{eq:slopes} is obtained in a region where the integral in \Eq{Boltzcom} is completely dominated by the UV. However, when $k$ approaches the cutoff, i.e. the end of the power-law behavior, the lower boundary of the integral becomes relevant and differences in the shape of the original spectrum start to matter.

To see this let us consider
a concrete example: 
$\phi \to N N \to \nu \nu \pi\pi$ with $N$ ($\nu, \pi$) being the right-handed neutrino (left-handed neutrino, pion). Here, for simplicity, we assume that the two body decay of $N$ is dominant, which happens in a certain parameter range~\cite{Gorbunov:2007ak}. Moreover, we neglect the masses of the decay products as well as neutrino flavor. 
As discussed in~\cite{Jaeckel:2020oet}, this process can carry the information from reheating to Earth without interacting with the plasma. 

We consider, again, three cases for the flux of $N$. 
In CASE A' $\f$ reheats the Universe. In CASE B' and C', {a subdominant $\f$ decays in the radiation dominated and matter dominated epoch, respectively. More precisely, we take the parameters for CASE A' corresponding to the CASE A set-up in the main text as 
\begin{eqnarray}
&&\!\!\!\!\{T_\f, m_\f, Br_{\f\to N{N} }\}
\\\nonumber
&&\quad\qquad\qquad\approx \{10\MEV, 1.2\times10^{16}\GEV,  14.4 \times 10^{-7}\}, 
\end{eqnarray}
for B' corresponding to B with parameters taken as
\beq
\{T_\f, m_\f, Br_{\f\to N{N} }\}\approx \{10\MEV, 1.2\times10^{16}\GEV,  0.7\},
\eeq
and $\rho_\f^{\rm ini}$ changed to be $20 T_\f^4,$ 
and for C' corresponding to C with 
\beq
\{ C_{\rm mat}, m_{\f } \}\approx \{2.7\times 10^{-61}\GEV^4,0.37\,{\rm PeV}\}. 
\eeq
In either case the other parameters are left unchanged.
Then we obtain the $\nu$ flux by using \eq{Boltzcom}, i.e. 
$ \frac{d^2 \F_\nu}{d \Omega d E_\nu }\simeq 1/2 \int{d \log {E_N} P[E_\nu, E_N] (E_N/E_\nu)^2 \frac{d^2 \F_N}{d \Omega  d E_N}}.$ We add a $1/2$ because on average half of the energy of $N$ is transferred into $\nu$. 
The resulting spectra are shown in Fig.~\ref{fig:ICECUBE}. (Note that we {display $E^2_{\nu} d^2 \F_\nu/d\Omega dE_{\nu}$} to match the presentation chosen by IceCube.)
The additional red solid line represents the reheating flux (CASE A) assuming a direct decay $\phi \to \nu \nu$ with $T_\f=10\MEV$, $m_\f\approx 9.5\times 10^{15}\GEV$ and $Br_{\f\to \nu\nu}=0.5\times 10^{-7}$.\footnote{\label{foot1}In fact, this is problematic since the early Universe is opaque to active neutrinos in standard cosmology. However, if $\f$ decays to dark fermions coupling to nucleons, we may get a similar spectrum. }

For an illustration of the experimental situation we also display current experimental data~\cite{Abbasi:2020jmh} from the IceCube experiment. For now all curves are consistent with the measurement (we have chosen the parameters such that they roughly fit the present data around the bump). A data point around $1$~PeV, i.e. just out of the range of the figure, is slightly above the curves. However, it may be easily explained by considering certain SM cosmic-rays e.g. \cite{Abbasi:2020jmh}. On the contrary the displayed bump may be difficult to explain within the SM since it is above the Waxman-Bahcall bound~\cite{Waxman:1998yy}. But, envisioning future improvements, it seems conceivable that the curves can be distinguished. 

We remark that the cascade decay spectrum of $\nu$ is a four body decay spectrum with a special phase space distribution. 
Therefore, even in the case of 
cascade decays or perhaps even multi-body decays, reheating may be probed in 
experiments by precisely measuring the spectra.

\begin{figure}[t!]
\begin{center}  
\includegraphics[width=75mm]{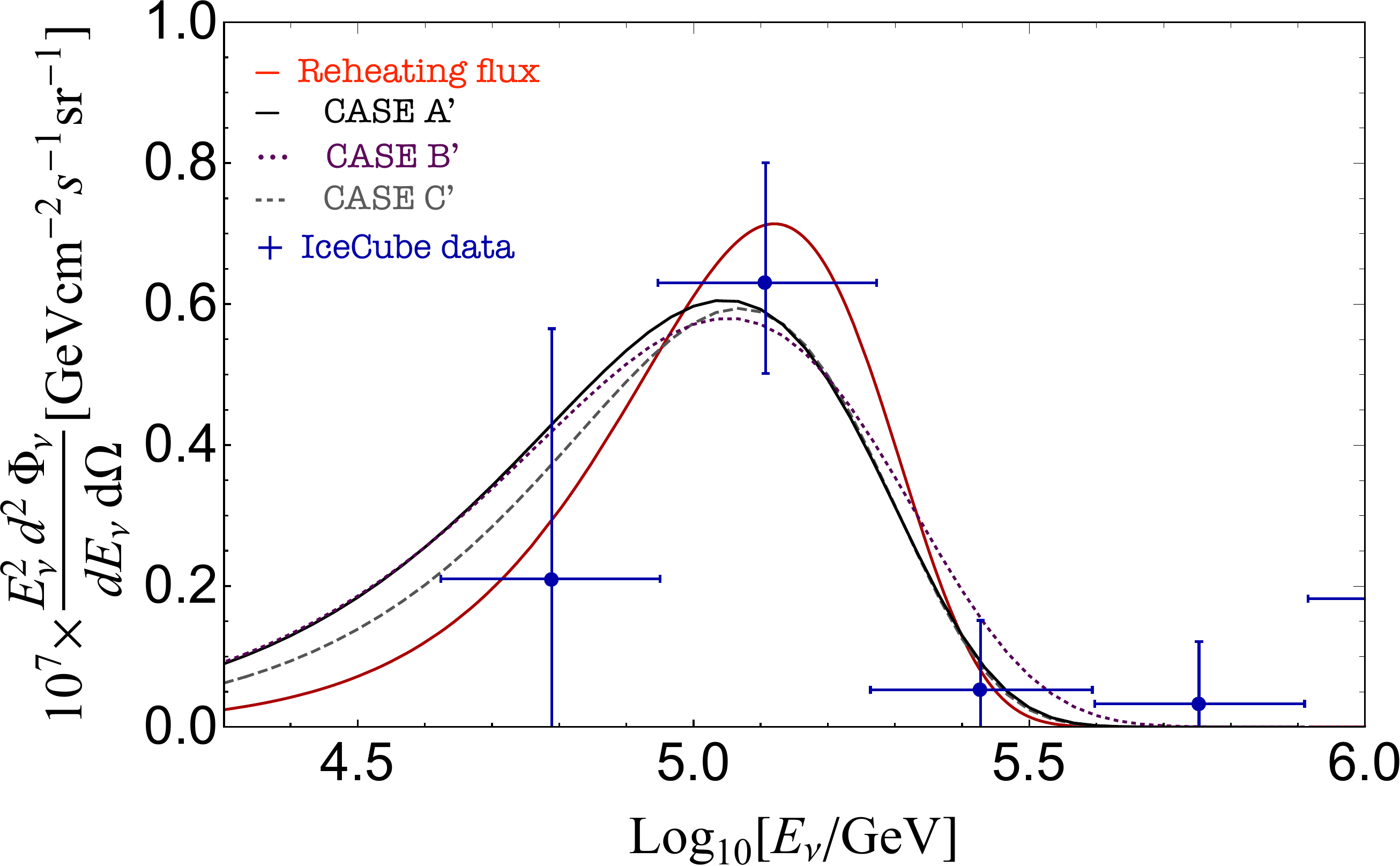}
\end{center}
\caption{Neutrino spectra from a cascade decay $\phi \to N N\to \n {\n} \pi\pi$. {Here CASE} A', B', and C' are shown in black solid, purple dotted, and gray dashed lines, respectively, representing the case that $\phi$ decays to reheat the Universe ($T_\phi=10\,$MeV), $\f$ decays in the radiation dominant era ($T_\phi=10\,$MeV), and decays in the matter dominant era ($z=100$). For comparison, we also {show a reheating} flux assuming $\phi \to \nu {\nu}$.\footref{foot1} 
IceCube 7.5 years' data (blue points) is taken from Ref.\,\cite{Abbasi:2020jmh}.
}
\label{fig:ICECUBE}
\end{figure}

\section{Statistics for measuring the reheating flux.}
\label{Appendix:C}
Let us roughly estimate what level of signal size we need in order to identify the reheating flux.
For simplicity of discussion, let us suppose that the  detector features only two energy bins close to the peak but in the power-law region of the flux.   
We can now try to discriminate the different power-laws of the flux that distinguish CASE A and B. 

We expect the number of events in a small energy bin to behave as (assuming that the sensitivity of the detector is energy independent over the relevant region),
\begin{equation}
    N\sim \frac{d\Phi}{dE} \Delta E\sim E^\gamma \Delta E,
\end{equation}
where we have also assumed a reasonably small bin size $\Delta E$ and that the flux scales as a power $\gamma$ of the energy.

Considering two energy bins $1$ and $2$ covering the energy interval $(E_0+2\Delta E,E_0+\Delta E)$ and $(E_0+\Delta E,E_0)$ we can now ask how many events we need in order to distinguish two different powers $\gamma$ such as $1/2$ and $1$, corresponding to CASE A and B, respectively.

The power can be estimated from
\begin{equation}
        \gamma\sim \frac{\log\left({N_{1}}/{N_{2}}\right)}{\log(1+\Delta E/E_{0})}.
\end{equation}
Treating the counting errors $\sim\sqrt{N_{1}}\sim \sqrt{N_{2}}$ in the two bins as statistically independent (adding the uncertainties in quadrature) we find,
\begin{equation}
    {\delta} \gamma\sim \frac{2}{\sqrt{N_1+N_2}(\Delta E/E_{0})}.
\end{equation}
For an $n\sigma$ distinction between two values of $\gamma$ differing by $\delta \gamma$ we therefore need,
\begin{equation}
    N_{1}+N_{2}\sim \frac{4 n^2}{\delta \gamma^2(\Delta E/E_{0})^2},
\end{equation}
events.

Using $\Delta E/E_{0}=0.25$ and $\delta\gamma=|1/2-1|=0.5$ (CASE A vs. CASE B) this yields roughly $\sim 1000$ events for a $2\sigma$ detection. Dropping the small $(\Delta E/E_{0})$ approximation this slightly increases to about $\sim 1400$.
That said, the procedure employed here is far from optimal. Choosing more and/or better separated bins this number can probably be decreased by an order of magnitude.

\bigskip
So far we have focused on the distinction between CASE A and CASE B.
Distinguishing CASE A from CASE C will be significantly more challenging
because the flux drops rapidly in the relevant region.
In this case, we will probably need to fit the flux by using a sizeable number of bins.  
Here, a very good measurement of the energy of the events, allowing for small bins as well high statistics will be required.

To put the above example into context let us just remark that in various future experiments, e.g. IAXO~\cite{Irastorza:2011gs,  Armengaud:2014gea, Armengaud:2019uso, Abeln:2020ywv}, IceCube~\cite{2009arXiv0907.2263W, Aartsen:2014gkd, Aartsen:2014njl, Aartsen:2020aqd} or DARWIN~\cite{Aalbers:2016jon}, the energy resolution for a new physics flux can be as small as $d \log_{10}{(E/\GEV)} \sim \O(0.1)$ or even better, for at least some energy range. In addition increased sensitivity, potentially by orders of magnitude, may then allow to collect the required statistics.

\bibliography{references}

\begin{thebibliography}{105}%
\makeatletter
\providecommand \@ifxundefined [1]{%
 \@ifx{#1\undefined}
}%
\providecommand \@ifnum [1]{%
 \ifnum #1\expandafter \@firstoftwo
 \else \expandafter \@secondoftwo
 \fi
}%
\providecommand \@ifx [1]{%
 \ifx #1\expandafter \@firstoftwo
 \else \expandafter \@secondoftwo
 \fi
}%
\providecommand \natexlab [1]{#1}%
\providecommand \enquote  [1]{``#1''}%
\providecommand \bibnamefont  [1]{#1}%
\providecommand \bibfnamefont [1]{#1}%
\providecommand \citenamefont [1]{#1}%
\providecommand \href@noop [0]{\@secondoftwo}%
\providecommand \href [0]{\begingroup \@sanitize@url \@href}%
\providecommand \@href[1]{\@@startlink{#1}\@@href}%
\providecommand \@@href[1]{\endgroup#1\@@endlink}%
\providecommand \@sanitize@url [0]{\catcode `\\12\catcode `\$12\catcode
  `\&12\catcode `\#12\catcode `\^12\catcode `\_12\catcode `\%12\relax}%
\providecommand \@@startlink[1]{}%
\providecommand \@@endlink[0]{}%
\providecommand \url  [0]{\begingroup\@sanitize@url \@url }%
\providecommand \@url [1]{\endgroup\@href {#1}{\urlprefix }}%
\providecommand \urlprefix  [0]{URL }%
\providecommand \Eprint [0]{\href }%
\providecommand \doibase [0]{http://dx.doi.org/}%
\providecommand \selectlanguage [0]{\@gobble}%
\providecommand \bibinfo  [0]{\@secondoftwo}%
\providecommand \bibfield  [0]{\@secondoftwo}%
\providecommand \translation [1]{[#1]}%
\providecommand \BibitemOpen [0]{}%
\providecommand \bibitemStop [0]{}%
\providecommand \bibitemNoStop [0]{.\EOS\space}%
\providecommand \EOS [0]{\spacefactor3000\relax}%
\providecommand \BibitemShut  [1]{\csname bibitem#1\endcsname}%
\let\auto@bib@innerbib\@empty
%</preamble>
\bibitem [{\citenamefont {Bassett}\ \emph {et~al.}(2006)\citenamefont
  {Bassett}, \citenamefont {Tsujikawa},\ and\ \citenamefont
  {Wands}}]{Bassett:2005xm}%
  \BibitemOpen
  \bibfield  {author} {\bibinfo {author} {\bibfnamefont {B.~A.}\ \bibnamefont
  {Bassett}}, \bibinfo {author} {\bibfnamefont {S.}~\bibnamefont {Tsujikawa}},
  \ and\ \bibinfo {author} {\bibfnamefont {D.}~\bibnamefont {Wands}},\ }\href
  {\doibase 10.1103/RevModPhys.78.537} {\bibfield  {journal} {\bibinfo
  {journal} {Rev. Mod. Phys.}\ }\textbf {\bibinfo {volume} {78}},\ \bibinfo
  {pages} {537} (\bibinfo {year} {2006})},\ \Eprint
  {http://arxiv.org/abs/astro-ph/0507632} {arXiv:astro-ph/0507632} \BibitemShut
  {NoStop}%
\bibitem [{\citenamefont {Kawasaki}\ \emph {et~al.}(1999)\citenamefont
  {Kawasaki}, \citenamefont {Kohri},\ and\ \citenamefont
  {Sugiyama}}]{Kawasaki:1999na}%
  \BibitemOpen
  \bibfield  {author} {\bibinfo {author} {\bibfnamefont {M.}~\bibnamefont
  {Kawasaki}}, \bibinfo {author} {\bibfnamefont {K.}~\bibnamefont {Kohri}}, \
  and\ \bibinfo {author} {\bibfnamefont {N.}~\bibnamefont {Sugiyama}},\ }\href
  {\doibase 10.1103/PhysRevLett.82.4168} {\bibfield  {journal} {\bibinfo
  {journal} {Phys. Rev. Lett.}\ }\textbf {\bibinfo {volume} {82}},\ \bibinfo
  {pages} {4168} (\bibinfo {year} {1999})},\ \Eprint
  {http://arxiv.org/abs/astro-ph/9811437} {arXiv:astro-ph/9811437 [astro-ph]}
  \BibitemShut {NoStop}%
%%CITATION = ASTRO-PH/9811437;%%
\bibitem [{\citenamefont {Kawasaki}\ \emph {et~al.}(2000)\citenamefont
  {Kawasaki}, \citenamefont {Kohri},\ and\ \citenamefont
  {Sugiyama}}]{Kawasaki:2000en}%
  \BibitemOpen
  \bibfield  {author} {\bibinfo {author} {\bibfnamefont {M.}~\bibnamefont
  {Kawasaki}}, \bibinfo {author} {\bibfnamefont {K.}~\bibnamefont {Kohri}}, \
  and\ \bibinfo {author} {\bibfnamefont {N.}~\bibnamefont {Sugiyama}},\ }\href
  {\doibase 10.1103/PhysRevD.62.023506} {\bibfield  {journal} {\bibinfo
  {journal} {Phys. Rev.}\ }\textbf {\bibinfo {volume} {D62}},\ \bibinfo {pages}
  {023506} (\bibinfo {year} {2000})},\ \Eprint
  {http://arxiv.org/abs/astro-ph/0002127} {arXiv:astro-ph/0002127 [astro-ph]}
  \BibitemShut {NoStop}%
%%CITATION = ASTRO-PH/0002127;%%
\bibitem [{\citenamefont {Hannestad}(2004)}]{Hannestad:2004px}%
  \BibitemOpen
  \bibfield  {author} {\bibinfo {author} {\bibfnamefont {S.}~\bibnamefont
  {Hannestad}},\ }\href {\doibase 10.1103/PhysRevD.70.043506} {\bibfield
  {journal} {\bibinfo  {journal} {Phys. Rev.}\ }\textbf {\bibinfo {volume}
  {D70}},\ \bibinfo {pages} {043506} (\bibinfo {year} {2004})},\ \Eprint
  {http://arxiv.org/abs/astro-ph/0403291} {arXiv:astro-ph/0403291 [astro-ph]}
  \BibitemShut {NoStop}%
%%CITATION = ASTRO-PH/0403291;%%
\bibitem [{\citenamefont {Ichikawa}\ \emph {et~al.}(2007)\citenamefont
  {Ichikawa}, \citenamefont {Kawasaki},\ and\ \citenamefont
  {Takahashi}}]{Ichikawa:2006vm}%
  \BibitemOpen
  \bibfield  {author} {\bibinfo {author} {\bibfnamefont {K.}~\bibnamefont
  {Ichikawa}}, \bibinfo {author} {\bibfnamefont {M.}~\bibnamefont {Kawasaki}},
  \ and\ \bibinfo {author} {\bibfnamefont {F.}~\bibnamefont {Takahashi}},\
  }\href {\doibase 10.1088/1475-7516/2007/05/007} {\bibfield  {journal}
  {\bibinfo  {journal} {JCAP}\ }\textbf {\bibinfo {volume} {0705}},\ \bibinfo
  {pages} {007} (\bibinfo {year} {2007})},\ \Eprint
  {http://arxiv.org/abs/astro-ph/0611784} {arXiv:astro-ph/0611784 [astro-ph]}
  \BibitemShut {NoStop}%
%%CITATION = ASTRO-PH/0611784;%%
\bibitem [{\citenamefont {De~Bernardis}\ \emph {et~al.}(2008)\citenamefont
  {De~Bernardis}, \citenamefont {Pagano},\ and\ \citenamefont
  {Melchiorri}}]{DeBernardis:2008zz}%
  \BibitemOpen
  \bibfield  {author} {\bibinfo {author} {\bibfnamefont {F.}~\bibnamefont
  {De~Bernardis}}, \bibinfo {author} {\bibfnamefont {L.}~\bibnamefont
  {Pagano}}, \ and\ \bibinfo {author} {\bibfnamefont {A.}~\bibnamefont
  {Melchiorri}},\ }\href {\doibase 10.1016/j.astropartphys.2008.09.005}
  {\bibfield  {journal} {\bibinfo  {journal} {Astropart. Phys.}\ }\textbf
  {\bibinfo {volume} {30}},\ \bibinfo {pages} {192} (\bibinfo {year}
  {2008})}\BibitemShut {NoStop}%
%%CITATION = APHYE,30,192;%%
\bibitem [{\citenamefont {de~Salas}\ \emph {et~al.}(2015)\citenamefont
  {de~Salas}, \citenamefont {Lattanzi}, \citenamefont {Mangano}, \citenamefont
  {Miele}, \citenamefont {Pastor},\ and\ \citenamefont
  {Pisanti}}]{deSalas:2015glj}%
  \BibitemOpen
  \bibfield  {author} {\bibinfo {author} {\bibfnamefont {P.~F.}\ \bibnamefont
  {de~Salas}}, \bibinfo {author} {\bibfnamefont {M.}~\bibnamefont {Lattanzi}},
  \bibinfo {author} {\bibfnamefont {G.}~\bibnamefont {Mangano}}, \bibinfo
  {author} {\bibfnamefont {G.}~\bibnamefont {Miele}}, \bibinfo {author}
  {\bibfnamefont {S.}~\bibnamefont {Pastor}}, \ and\ \bibinfo {author}
  {\bibfnamefont {O.}~\bibnamefont {Pisanti}},\ }\href {\doibase
  10.1103/PhysRevD.92.123534} {\bibfield  {journal} {\bibinfo  {journal} {Phys.
  Rev.}\ }\textbf {\bibinfo {volume} {D92}},\ \bibinfo {pages} {123534}
  (\bibinfo {year} {2015})},\ \Eprint {http://arxiv.org/abs/1511.00672}
  {arXiv:1511.00672 [astro-ph.CO]} \BibitemShut {NoStop}%
%%CITATION = ARXIV:1511.00672;%%
\bibitem [{\citenamefont {Hufnagel}\ \emph {et~al.}(2018)\citenamefont
  {Hufnagel}, \citenamefont {Schmidt-Hoberg},\ and\ \citenamefont
  {Wild}}]{Hufnagel:2018bjp}%
  \BibitemOpen
  \bibfield  {author} {\bibinfo {author} {\bibfnamefont {M.}~\bibnamefont
  {Hufnagel}}, \bibinfo {author} {\bibfnamefont {K.}~\bibnamefont
  {Schmidt-Hoberg}}, \ and\ \bibinfo {author} {\bibfnamefont {S.}~\bibnamefont
  {Wild}},\ }\href {\doibase 10.1088/1475-7516/2018/11/032} {\bibfield
  {journal} {\bibinfo  {journal} {JCAP}\ }\textbf {\bibinfo {volume} {11}},\
  \bibinfo {pages} {032} (\bibinfo {year} {2018})},\ \Eprint
  {http://arxiv.org/abs/1808.09324} {arXiv:1808.09324 [hep-ph]} \BibitemShut
  {NoStop}%
\bibitem [{\citenamefont {Hasegawa}\ \emph {et~al.}(2019)\citenamefont
  {Hasegawa}, \citenamefont {Hiroshima}, \citenamefont {Kohri}, \citenamefont
  {Hansen}, \citenamefont {Tram},\ and\ \citenamefont
  {Hannestad}}]{Hasegawa:2019jsa}%
  \BibitemOpen
  \bibfield  {author} {\bibinfo {author} {\bibfnamefont {T.}~\bibnamefont
  {Hasegawa}}, \bibinfo {author} {\bibfnamefont {N.}~\bibnamefont {Hiroshima}},
  \bibinfo {author} {\bibfnamefont {K.}~\bibnamefont {Kohri}}, \bibinfo
  {author} {\bibfnamefont {R.~S.~L.}\ \bibnamefont {Hansen}}, \bibinfo {author}
  {\bibfnamefont {T.}~\bibnamefont {Tram}}, \ and\ \bibinfo {author}
  {\bibfnamefont {S.}~\bibnamefont {Hannestad}},\ }\href {\doibase
  10.1088/1475-7516/2019/12/012} {\bibfield  {journal} {\bibinfo  {journal}
  {JCAP}\ }\textbf {\bibinfo {volume} {1912}},\ \bibinfo {pages} {012}
  (\bibinfo {year} {2019})},\ \Eprint {http://arxiv.org/abs/1908.10189}
  {arXiv:1908.10189 [hep-ph]} \BibitemShut {NoStop}%
%%CITATION = ARXIV:1908.10189;%%
\bibitem [{\citenamefont {Kawasaki}\ \emph {et~al.}(2020)\citenamefont
  {Kawasaki}, \citenamefont {Kohri}, \citenamefont {Moroi}, \citenamefont
  {Murai},\ and\ \citenamefont {Murayama}}]{Kawasaki:2020qxm}%
  \BibitemOpen
  \bibfield  {author} {\bibinfo {author} {\bibfnamefont {M.}~\bibnamefont
  {Kawasaki}}, \bibinfo {author} {\bibfnamefont {K.}~\bibnamefont {Kohri}},
  \bibinfo {author} {\bibfnamefont {T.}~\bibnamefont {Moroi}}, \bibinfo
  {author} {\bibfnamefont {K.}~\bibnamefont {Murai}}, \ and\ \bibinfo {author}
  {\bibfnamefont {H.}~\bibnamefont {Murayama}},\ }\href {\doibase
  10.1088/1475-7516/2020/12/048} {\bibfield  {journal} {\bibinfo  {journal}
  {JCAP}\ }\textbf {\bibinfo {volume} {12}},\ \bibinfo {pages} {048} (\bibinfo
  {year} {2020})},\ \Eprint {http://arxiv.org/abs/2006.14803} {arXiv:2006.14803
  [hep-ph]} \BibitemShut {NoStop}%
\bibitem [{\citenamefont {Depta}\ \emph {et~al.}(2020)\citenamefont {Depta},
  \citenamefont {Hufnagel},\ and\ \citenamefont
  {Schmidt-Hoberg}}]{Depta:2020zbh}%
  \BibitemOpen
  \bibfield  {author} {\bibinfo {author} {\bibfnamefont {P.~F.}\ \bibnamefont
  {Depta}}, \bibinfo {author} {\bibfnamefont {M.}~\bibnamefont {Hufnagel}}, \
  and\ \bibinfo {author} {\bibfnamefont {K.}~\bibnamefont {Schmidt-Hoberg}},\
  }\href@noop {} {\  (\bibinfo {year} {2020})},\ \Eprint
  {http://arxiv.org/abs/2011.06519} {arXiv:2011.06519 [hep-ph]} \BibitemShut
  {NoStop}%
\bibitem [{\citenamefont {Akrami}\ \emph {et~al.}(2018)\citenamefont {Akrami}
  \emph {et~al.}}]{Akrami:2018odb}%
  \BibitemOpen
  \bibfield  {author} {\bibinfo {author} {\bibfnamefont {Y.}~\bibnamefont
  {Akrami}} \emph {et~al.} (\bibinfo {collaboration} {Planck}),\ }\href@noop {}
  {\  (\bibinfo {year} {2018})},\ \Eprint {http://arxiv.org/abs/1807.06211}
  {arXiv:1807.06211 [astro-ph.CO]} \BibitemShut {NoStop}%
%%CITATION = ARXIV:1807.06211;%%
\bibitem [{\citenamefont {Tashiro}\ \emph {et~al.}(2004)\citenamefont
  {Tashiro}, \citenamefont {Chiba},\ and\ \citenamefont
  {Sasaki}}]{Tashiro:2003qp}%
  \BibitemOpen
  \bibfield  {author} {\bibinfo {author} {\bibfnamefont {H.}~\bibnamefont
  {Tashiro}}, \bibinfo {author} {\bibfnamefont {T.}~\bibnamefont {Chiba}}, \
  and\ \bibinfo {author} {\bibfnamefont {M.}~\bibnamefont {Sasaki}},\ }\href
  {\doibase 10.1088/0264-9381/21/7/004} {\bibfield  {journal} {\bibinfo
  {journal} {Class. Quant. Grav.}\ }\textbf {\bibinfo {volume} {21}},\ \bibinfo
  {pages} {1761} (\bibinfo {year} {2004})},\ \Eprint
  {http://arxiv.org/abs/gr-qc/0307068} {arXiv:gr-qc/0307068} \BibitemShut
  {NoStop}%
\bibitem [{\citenamefont {Easther}\ and\ \citenamefont
  {Lim}(2006)}]{Easther:2006gt}%
  \BibitemOpen
  \bibfield  {author} {\bibinfo {author} {\bibfnamefont {R.}~\bibnamefont
  {Easther}}\ and\ \bibinfo {author} {\bibfnamefont {E.~A.}\ \bibnamefont
  {Lim}},\ }\href {\doibase 10.1088/1475-7516/2006/04/010} {\bibfield
  {journal} {\bibinfo  {journal} {JCAP}\ }\textbf {\bibinfo {volume} {04}},\
  \bibinfo {pages} {010} (\bibinfo {year} {2006})},\ \Eprint
  {http://arxiv.org/abs/astro-ph/0601617} {arXiv:astro-ph/0601617} \BibitemShut
  {NoStop}%
\bibitem [{\citenamefont {Garcia-Bellido}\ \emph {et~al.}(2008)\citenamefont
  {Garcia-Bellido}, \citenamefont {Figueroa},\ and\ \citenamefont
  {Sastre}}]{GarciaBellido:2007af}%
  \BibitemOpen
  \bibfield  {author} {\bibinfo {author} {\bibfnamefont {J.}~\bibnamefont
  {Garcia-Bellido}}, \bibinfo {author} {\bibfnamefont {D.~G.}\ \bibnamefont
  {Figueroa}}, \ and\ \bibinfo {author} {\bibfnamefont {A.}~\bibnamefont
  {Sastre}},\ }\href {\doibase 10.1103/PhysRevD.77.043517} {\bibfield
  {journal} {\bibinfo  {journal} {Phys. Rev. D}\ }\textbf {\bibinfo {volume}
  {77}},\ \bibinfo {pages} {043517} (\bibinfo {year} {2008})},\ \Eprint
  {http://arxiv.org/abs/0707.0839} {arXiv:0707.0839 [hep-ph]} \BibitemShut
  {NoStop}%
\bibitem [{\citenamefont {Dufaux}\ \emph {et~al.}(2007)\citenamefont {Dufaux},
  \citenamefont {Bergman}, \citenamefont {Felder}, \citenamefont {Kofman},\
  and\ \citenamefont {Uzan}}]{Dufaux:2007pt}%
  \BibitemOpen
  \bibfield  {author} {\bibinfo {author} {\bibfnamefont {J.~F.}\ \bibnamefont
  {Dufaux}}, \bibinfo {author} {\bibfnamefont {A.}~\bibnamefont {Bergman}},
  \bibinfo {author} {\bibfnamefont {G.~N.}\ \bibnamefont {Felder}}, \bibinfo
  {author} {\bibfnamefont {L.}~\bibnamefont {Kofman}}, \ and\ \bibinfo {author}
  {\bibfnamefont {J.-P.}\ \bibnamefont {Uzan}},\ }\href {\doibase
  10.1103/PhysRevD.76.123517} {\bibfield  {journal} {\bibinfo  {journal} {Phys.
  Rev. D}\ }\textbf {\bibinfo {volume} {76}},\ \bibinfo {pages} {123517}
  (\bibinfo {year} {2007})},\ \Eprint {http://arxiv.org/abs/0707.0875}
  {arXiv:0707.0875 [astro-ph]} \BibitemShut {NoStop}%
\bibitem [{\citenamefont {Huang}(2011)}]{Huang:2011gf}%
  \BibitemOpen
  \bibfield  {author} {\bibinfo {author} {\bibfnamefont {Z.}~\bibnamefont
  {Huang}},\ }\href {\doibase 10.1103/PhysRevD.83.123509} {\bibfield  {journal}
  {\bibinfo  {journal} {Phys. Rev. D}\ }\textbf {\bibinfo {volume} {83}},\
  \bibinfo {pages} {123509} (\bibinfo {year} {2011})},\ \Eprint
  {http://arxiv.org/abs/1102.0227} {arXiv:1102.0227 [astro-ph.CO]} \BibitemShut
  {NoStop}%
\bibitem [{\citenamefont {Hebecker}\ \emph {et~al.}(2016)\citenamefont
  {Hebecker}, \citenamefont {Jaeckel}, \citenamefont {Rompineve},\ and\
  \citenamefont {Witkowski}}]{Hebecker:2016vbl}%
  \BibitemOpen
  \bibfield  {author} {\bibinfo {author} {\bibfnamefont {A.}~\bibnamefont
  {Hebecker}}, \bibinfo {author} {\bibfnamefont {J.}~\bibnamefont {Jaeckel}},
  \bibinfo {author} {\bibfnamefont {F.}~\bibnamefont {Rompineve}}, \ and\
  \bibinfo {author} {\bibfnamefont {L.~T.}\ \bibnamefont {Witkowski}},\ }\href
  {\doibase 10.1088/1475-7516/2016/11/003} {\bibfield  {journal} {\bibinfo
  {journal} {JCAP}\ }\textbf {\bibinfo {volume} {11}},\ \bibinfo {pages} {003}
  (\bibinfo {year} {2016})},\ \Eprint {http://arxiv.org/abs/1606.07812}
  {arXiv:1606.07812 [hep-ph]} \BibitemShut {NoStop}%
\bibitem [{\citenamefont {Amin}\ \emph
  {et~al.}(2019{\natexlab{a}})\citenamefont {Amin}, \citenamefont {Fan},
  \citenamefont {Lozanov},\ and\ \citenamefont {Reece}}]{Amin:2018kkg}%
  \BibitemOpen
  \bibfield  {author} {\bibinfo {author} {\bibfnamefont {M.~A.}\ \bibnamefont
  {Amin}}, \bibinfo {author} {\bibfnamefont {J.}~\bibnamefont {Fan}}, \bibinfo
  {author} {\bibfnamefont {K.~D.}\ \bibnamefont {Lozanov}}, \ and\ \bibinfo
  {author} {\bibfnamefont {M.}~\bibnamefont {Reece}},\ }\href {\doibase
  10.1103/PhysRevD.99.035008} {\bibfield  {journal} {\bibinfo  {journal} {Phys.
  Rev. D}\ }\textbf {\bibinfo {volume} {99}},\ \bibinfo {pages} {035008}
  (\bibinfo {year} {2019}{\natexlab{a}})},\ \Eprint
  {http://arxiv.org/abs/1802.00444} {arXiv:1802.00444 [hep-ph]} \BibitemShut
  {NoStop}%
\bibitem [{\citenamefont {Adshead}\ \emph {et~al.}(2018)\citenamefont
  {Adshead}, \citenamefont {Giblin},\ and\ \citenamefont
  {Weiner}}]{Adshead:2018doq}%
  \BibitemOpen
  \bibfield  {author} {\bibinfo {author} {\bibfnamefont {P.}~\bibnamefont
  {Adshead}}, \bibinfo {author} {\bibfnamefont {J.~T.}\ \bibnamefont {Giblin}},
  \ and\ \bibinfo {author} {\bibfnamefont {Z.~J.}\ \bibnamefont {Weiner}},\
  }\href {\doibase 10.1103/PhysRevD.98.043525} {\bibfield  {journal} {\bibinfo
  {journal} {Phys. Rev. D}\ }\textbf {\bibinfo {volume} {98}},\ \bibinfo
  {pages} {043525} (\bibinfo {year} {2018})},\ \Eprint
  {http://arxiv.org/abs/1805.04550} {arXiv:1805.04550 [astro-ph.CO]}
  \BibitemShut {NoStop}%
\bibitem [{\citenamefont {Kitajima}\ \emph
  {et~al.}(2018{\natexlab{a}})\citenamefont {Kitajima}, \citenamefont {Soda},\
  and\ \citenamefont {Urakawa}}]{Kitajima:2018zco}%
  \BibitemOpen
  \bibfield  {author} {\bibinfo {author} {\bibfnamefont {N.}~\bibnamefont
  {Kitajima}}, \bibinfo {author} {\bibfnamefont {J.}~\bibnamefont {Soda}}, \
  and\ \bibinfo {author} {\bibfnamefont {Y.}~\bibnamefont {Urakawa}},\ }\href
  {\doibase 10.1088/1475-7516/2018/10/008} {\bibfield  {journal} {\bibinfo
  {journal} {JCAP}\ }\textbf {\bibinfo {volume} {10}},\ \bibinfo {pages} {008}
  (\bibinfo {year} {2018}{\natexlab{a}})},\ \Eprint
  {http://arxiv.org/abs/1807.07037} {arXiv:1807.07037 [astro-ph.CO]}
  \BibitemShut {NoStop}%
\bibitem [{\citenamefont {Nakayama}\ and\ \citenamefont
  {Tang}(2019)}]{Nakayama:2018ptw}%
  \BibitemOpen
  \bibfield  {author} {\bibinfo {author} {\bibfnamefont {K.}~\bibnamefont
  {Nakayama}}\ and\ \bibinfo {author} {\bibfnamefont {Y.}~\bibnamefont
  {Tang}},\ }\href {\doibase 10.1016/j.physletb.2018.11.023} {\bibfield
  {journal} {\bibinfo  {journal} {Phys. Lett. B}\ }\textbf {\bibinfo {volume}
  {788}},\ \bibinfo {pages} {341} (\bibinfo {year} {2019})},\ \Eprint
  {http://arxiv.org/abs/1810.04975} {arXiv:1810.04975 [hep-ph]} \BibitemShut
  {NoStop}%
\bibitem [{\citenamefont {Lozanov}\ and\ \citenamefont
  {Amin}(2019)}]{Lozanov:2019ylm}%
  \BibitemOpen
  \bibfield  {author} {\bibinfo {author} {\bibfnamefont {K.~D.}\ \bibnamefont
  {Lozanov}}\ and\ \bibinfo {author} {\bibfnamefont {M.~A.}\ \bibnamefont
  {Amin}},\ }\href {\doibase 10.1103/PhysRevD.99.123504} {\bibfield  {journal}
  {\bibinfo  {journal} {Phys. Rev. D}\ }\textbf {\bibinfo {volume} {99}},\
  \bibinfo {pages} {123504} (\bibinfo {year} {2019})},\ \Eprint
  {http://arxiv.org/abs/1902.06736} {arXiv:1902.06736 [astro-ph.CO]}
  \BibitemShut {NoStop}%
\bibitem [{\citenamefont {Sang}\ and\ \citenamefont
  {Huang}(2019)}]{Sang:2019ndv}%
  \BibitemOpen
  \bibfield  {author} {\bibinfo {author} {\bibfnamefont {Y.}~\bibnamefont
  {Sang}}\ and\ \bibinfo {author} {\bibfnamefont {Q.-G.}\ \bibnamefont
  {Huang}},\ }\href {\doibase 10.1103/PhysRevD.100.063516} {\bibfield
  {journal} {\bibinfo  {journal} {Phys. Rev. D}\ }\textbf {\bibinfo {volume}
  {100}},\ \bibinfo {pages} {063516} (\bibinfo {year} {2019})},\ \Eprint
  {http://arxiv.org/abs/1905.00371} {arXiv:1905.00371 [astro-ph.CO]}
  \BibitemShut {NoStop}%
\bibitem [{\citenamefont {Adshead}\ \emph
  {et~al.}(2020{\natexlab{a}})\citenamefont {Adshead}, \citenamefont {Giblin},
  \citenamefont {Pieroni},\ and\ \citenamefont {Weiner}}]{Adshead:2019lbr}%
  \BibitemOpen
  \bibfield  {author} {\bibinfo {author} {\bibfnamefont {P.}~\bibnamefont
  {Adshead}}, \bibinfo {author} {\bibfnamefont {J.~T.}\ \bibnamefont {Giblin}},
  \bibinfo {author} {\bibfnamefont {M.}~\bibnamefont {Pieroni}}, \ and\
  \bibinfo {author} {\bibfnamefont {Z.~J.}\ \bibnamefont {Weiner}},\ }\href
  {\doibase 10.1103/PhysRevD.101.083534} {\bibfield  {journal} {\bibinfo
  {journal} {Phys. Rev. D}\ }\textbf {\bibinfo {volume} {101}},\ \bibinfo
  {pages} {083534} (\bibinfo {year} {2020}{\natexlab{a}})},\ \Eprint
  {http://arxiv.org/abs/1909.12842} {arXiv:1909.12842 [astro-ph.CO]}
  \BibitemShut {NoStop}%
\bibitem [{\citenamefont {Adshead}\ \emph
  {et~al.}(2020{\natexlab{b}})\citenamefont {Adshead}, \citenamefont {Giblin},
  \citenamefont {Pieroni},\ and\ \citenamefont {Weiner}}]{Adshead:2019igv}%
  \BibitemOpen
  \bibfield  {author} {\bibinfo {author} {\bibfnamefont {P.}~\bibnamefont
  {Adshead}}, \bibinfo {author} {\bibfnamefont {J.~T.}\ \bibnamefont {Giblin}},
  \bibinfo {author} {\bibfnamefont {M.}~\bibnamefont {Pieroni}}, \ and\
  \bibinfo {author} {\bibfnamefont {Z.~J.}\ \bibnamefont {Weiner}},\ }\href
  {\doibase 10.1103/PhysRevLett.124.171301} {\bibfield  {journal} {\bibinfo
  {journal} {Phys. Rev. Lett.}\ }\textbf {\bibinfo {volume} {124}},\ \bibinfo
  {pages} {171301} (\bibinfo {year} {2020}{\natexlab{b}})},\ \Eprint
  {http://arxiv.org/abs/1909.12843} {arXiv:1909.12843 [astro-ph.CO]}
  \BibitemShut {NoStop}%
\bibitem [{\citenamefont {Domcke}\ \emph {et~al.}(2020)\citenamefont {Domcke},
  \citenamefont {Jinno},\ and\ \citenamefont {Rubira}}]{Domcke:2020xmn}%
  \BibitemOpen
  \bibfield  {author} {\bibinfo {author} {\bibfnamefont {V.}~\bibnamefont
  {Domcke}}, \bibinfo {author} {\bibfnamefont {R.}~\bibnamefont {Jinno}}, \
  and\ \bibinfo {author} {\bibfnamefont {H.}~\bibnamefont {Rubira}},\ }\href
  {\doibase 10.1088/1475-7516/2020/06/046} {\bibfield  {journal} {\bibinfo
  {journal} {JCAP}\ }\textbf {\bibinfo {volume} {06}},\ \bibinfo {pages} {046}
  (\bibinfo {year} {2020})},\ \Eprint {http://arxiv.org/abs/2002.11083}
  {arXiv:2002.11083 [astro-ph.CO]} \BibitemShut {NoStop}%
\bibitem [{\citenamefont {Cicoli}\ \emph {et~al.}(2013)\citenamefont {Cicoli},
  \citenamefont {Conlon},\ and\ \citenamefont {Quevedo}}]{Cicoli:2012aq}%
  \BibitemOpen
  \bibfield  {author} {\bibinfo {author} {\bibfnamefont {M.}~\bibnamefont
  {Cicoli}}, \bibinfo {author} {\bibfnamefont {J.~P.}\ \bibnamefont {Conlon}},
  \ and\ \bibinfo {author} {\bibfnamefont {F.}~\bibnamefont {Quevedo}},\ }\href
  {\doibase 10.1103/PhysRevD.87.043520} {\bibfield  {journal} {\bibinfo
  {journal} {Phys. Rev. D}\ }\textbf {\bibinfo {volume} {87}},\ \bibinfo
  {pages} {043520} (\bibinfo {year} {2013})},\ \Eprint
  {http://arxiv.org/abs/1208.3562} {arXiv:1208.3562 [hep-ph]} \BibitemShut
  {NoStop}%
\bibitem [{\citenamefont {Higaki}\ and\ \citenamefont
  {Takahashi}(2012)}]{Higaki:2012ar}%
  \BibitemOpen
  \bibfield  {author} {\bibinfo {author} {\bibfnamefont {T.}~\bibnamefont
  {Higaki}}\ and\ \bibinfo {author} {\bibfnamefont {F.}~\bibnamefont
  {Takahashi}},\ }\href {\doibase 10.1007/JHEP11(2012)125} {\bibfield
  {journal} {\bibinfo  {journal} {JHEP}\ }\textbf {\bibinfo {volume} {11}},\
  \bibinfo {pages} {125} (\bibinfo {year} {2012})},\ \Eprint
  {http://arxiv.org/abs/1208.3563} {arXiv:1208.3563 [hep-ph]} \BibitemShut
  {NoStop}%
\bibitem [{\citenamefont {Conlon}\ and\ \citenamefont
  {Marsh}(2013)}]{Conlon:2013isa}%
  \BibitemOpen
  \bibfield  {author} {\bibinfo {author} {\bibfnamefont {J.~P.}\ \bibnamefont
  {Conlon}}\ and\ \bibinfo {author} {\bibfnamefont {M.~C.~D.}\ \bibnamefont
  {Marsh}},\ }\href {\doibase 10.1007/JHEP10(2013)214} {\bibfield  {journal}
  {\bibinfo  {journal} {JHEP}\ }\textbf {\bibinfo {volume} {10}},\ \bibinfo
  {pages} {214} (\bibinfo {year} {2013})},\ \Eprint
  {http://arxiv.org/abs/1304.1804} {arXiv:1304.1804 [hep-ph]} \BibitemShut
  {NoStop}%
\bibitem [{\citenamefont {Hebecker}\ \emph {et~al.}(2014)\citenamefont
  {Hebecker}, \citenamefont {Mangat}, \citenamefont {Rompineve},\ and\
  \citenamefont {Witkowski}}]{Hebecker:2014gka}%
  \BibitemOpen
  \bibfield  {author} {\bibinfo {author} {\bibfnamefont {A.}~\bibnamefont
  {Hebecker}}, \bibinfo {author} {\bibfnamefont {P.}~\bibnamefont {Mangat}},
  \bibinfo {author} {\bibfnamefont {F.}~\bibnamefont {Rompineve}}, \ and\
  \bibinfo {author} {\bibfnamefont {L.~T.}\ \bibnamefont {Witkowski}},\ }\href
  {\doibase 10.1007/JHEP09(2014)140} {\bibfield  {journal} {\bibinfo  {journal}
  {JHEP}\ }\textbf {\bibinfo {volume} {09}},\ \bibinfo {pages} {140} (\bibinfo
  {year} {2014})},\ \Eprint {http://arxiv.org/abs/1403.6810} {arXiv:1403.6810
  [hep-ph]} \BibitemShut {NoStop}%
\bibitem [{\citenamefont {Jaeckel}\ and\ \citenamefont
  {Yin}(2020)}]{Jaeckel:2020oet}%
  \BibitemOpen
  \bibfield  {author} {\bibinfo {author} {\bibfnamefont {J.}~\bibnamefont
  {Jaeckel}}\ and\ \bibinfo {author} {\bibfnamefont {W.}~\bibnamefont {Yin}},\
  }\href@noop {} {\  (\bibinfo {year} {2020})},\ \Eprint
  {http://arxiv.org/abs/2007.15006} {arXiv:2007.15006 [hep-ph]} \BibitemShut
  {NoStop}%
\bibitem [{\citenamefont {Armengaud}\ \emph {et~al.}(2019)\citenamefont
  {Armengaud} \emph {et~al.}}]{Armengaud:2019uso}%
  \BibitemOpen
  \bibfield  {author} {\bibinfo {author} {\bibfnamefont {E.}~\bibnamefont
  {Armengaud}} \emph {et~al.} (\bibinfo {collaboration} {IAXO}),\ }\href
  {\doibase 10.1088/1475-7516/2019/06/047} {\bibfield  {journal} {\bibinfo
  {journal} {JCAP}\ }\textbf {\bibinfo {volume} {06}},\ \bibinfo {pages} {047}
  (\bibinfo {year} {2019})},\ \Eprint {http://arxiv.org/abs/1904.09155}
  {arXiv:1904.09155 [hep-ph]} \BibitemShut {NoStop}%
\bibitem [{\citenamefont {Ema}\ \emph {et~al.}(2014{\natexlab{a}})\citenamefont
  {Ema}, \citenamefont {Jinno},\ and\ \citenamefont {Moroi}}]{Ema:2013nda}%
  \BibitemOpen
  \bibfield  {author} {\bibinfo {author} {\bibfnamefont {Y.}~\bibnamefont
  {Ema}}, \bibinfo {author} {\bibfnamefont {R.}~\bibnamefont {Jinno}}, \ and\
  \bibinfo {author} {\bibfnamefont {T.}~\bibnamefont {Moroi}},\ }\href
  {\doibase 10.1016/j.physletb.2014.04.021} {\bibfield  {journal} {\bibinfo
  {journal} {Phys. Lett.}\ }\textbf {\bibinfo {volume} {B733}},\ \bibinfo
  {pages} {120} (\bibinfo {year} {2014}{\natexlab{a}})},\ \Eprint
  {http://arxiv.org/abs/1312.3501} {arXiv:1312.3501 [hep-ph]} \BibitemShut
  {NoStop}%
%%CITATION = ARXIV:1312.3501;%%
\bibitem [{\citenamefont {Ema}\ \emph {et~al.}(2014{\natexlab{b}})\citenamefont
  {Ema}, \citenamefont {Jinno},\ and\ \citenamefont {Moroi}}]{Ema:2014ufa}%
  \BibitemOpen
  \bibfield  {author} {\bibinfo {author} {\bibfnamefont {Y.}~\bibnamefont
  {Ema}}, \bibinfo {author} {\bibfnamefont {R.}~\bibnamefont {Jinno}}, \ and\
  \bibinfo {author} {\bibfnamefont {T.}~\bibnamefont {Moroi}},\ }\href
  {\doibase 10.1007/JHEP10(2014)150} {\bibfield  {journal} {\bibinfo  {journal}
  {JHEP}\ }\textbf {\bibinfo {volume} {10}},\ \bibinfo {pages} {150} (\bibinfo
  {year} {2014}{\natexlab{b}})},\ \Eprint {http://arxiv.org/abs/1408.1745}
  {arXiv:1408.1745 [hep-ph]} \BibitemShut {NoStop}%
%%CITATION = ARXIV:1408.1745;%%
\bibitem [{\citenamefont {Traschen}\ and\ \citenamefont
  {Brandenberger}(1990)}]{Traschen:1990sw}%
  \BibitemOpen
  \bibfield  {author} {\bibinfo {author} {\bibfnamefont {J.~H.}\ \bibnamefont
  {Traschen}}\ and\ \bibinfo {author} {\bibfnamefont {R.~H.}\ \bibnamefont
  {Brandenberger}},\ }\href {\doibase 10.1103/PhysRevD.42.2491} {\bibfield
  {journal} {\bibinfo  {journal} {Phys. Rev. D}\ }\textbf {\bibinfo {volume}
  {42}},\ \bibinfo {pages} {2491} (\bibinfo {year} {1990})}\BibitemShut
  {NoStop}%
\bibitem [{\citenamefont {Kofman}\ \emph {et~al.}(1994)\citenamefont {Kofman},
  \citenamefont {Linde},\ and\ \citenamefont {Starobinsky}}]{Kofman:1994rk}%
  \BibitemOpen
  \bibfield  {author} {\bibinfo {author} {\bibfnamefont {L.}~\bibnamefont
  {Kofman}}, \bibinfo {author} {\bibfnamefont {A.~D.}\ \bibnamefont {Linde}}, \
  and\ \bibinfo {author} {\bibfnamefont {A.~A.}\ \bibnamefont {Starobinsky}},\
  }\href {\doibase 10.1103/PhysRevLett.73.3195} {\bibfield  {journal} {\bibinfo
   {journal} {Phys. Rev. Lett.}\ }\textbf {\bibinfo {volume} {73}},\ \bibinfo
  {pages} {3195} (\bibinfo {year} {1994})},\ \Eprint
  {http://arxiv.org/abs/hep-th/9405187} {arXiv:hep-th/9405187} \BibitemShut
  {NoStop}%
\bibitem [{\citenamefont {Shtanov}\ \emph {et~al.}(1995)\citenamefont
  {Shtanov}, \citenamefont {Traschen},\ and\ \citenamefont
  {Brandenberger}}]{Shtanov:1994ce}%
  \BibitemOpen
  \bibfield  {author} {\bibinfo {author} {\bibfnamefont {Y.}~\bibnamefont
  {Shtanov}}, \bibinfo {author} {\bibfnamefont {J.~H.}\ \bibnamefont
  {Traschen}}, \ and\ \bibinfo {author} {\bibfnamefont {R.~H.}\ \bibnamefont
  {Brandenberger}},\ }\href {\doibase 10.1103/PhysRevD.51.5438} {\bibfield
  {journal} {\bibinfo  {journal} {Phys. Rev. D}\ }\textbf {\bibinfo {volume}
  {51}},\ \bibinfo {pages} {5438} (\bibinfo {year} {1995})},\ \Eprint
  {http://arxiv.org/abs/hep-ph/9407247} {arXiv:hep-ph/9407247} \BibitemShut
  {NoStop}%
\bibitem [{\citenamefont {Yoshimura}(1995)}]{Yoshimura:1995gc}%
  \BibitemOpen
  \bibfield  {author} {\bibinfo {author} {\bibfnamefont {M.}~\bibnamefont
  {Yoshimura}},\ }\href {\doibase 10.1143/PTP.94.873} {\bibfield  {journal}
  {\bibinfo  {journal} {Prog. Theor. Phys.}\ }\textbf {\bibinfo {volume}
  {94}},\ \bibinfo {pages} {873} (\bibinfo {year} {1995})},\ \Eprint
  {http://arxiv.org/abs/hep-th/9506176} {arXiv:hep-th/9506176} \BibitemShut
  {NoStop}%
\bibitem [{\citenamefont {Kasuya}\ and\ \citenamefont
  {Kawasaki}(1996)}]{Kasuya:1996np}%
  \BibitemOpen
  \bibfield  {author} {\bibinfo {author} {\bibfnamefont {S.}~\bibnamefont
  {Kasuya}}\ and\ \bibinfo {author} {\bibfnamefont {M.}~\bibnamefont
  {Kawasaki}},\ }\href {\doibase 10.1016/S0370-2693(96)01216-6} {\bibfield
  {journal} {\bibinfo  {journal} {Phys. Lett. B}\ }\textbf {\bibinfo {volume}
  {388}},\ \bibinfo {pages} {686} (\bibinfo {year} {1996})},\ \Eprint
  {http://arxiv.org/abs/hep-ph/9603317} {arXiv:hep-ph/9603317} \BibitemShut
  {NoStop}%
\bibitem [{\citenamefont {Kofman}\ \emph {et~al.}(1997)\citenamefont {Kofman},
  \citenamefont {Linde},\ and\ \citenamefont {Starobinsky}}]{Kofman:1997yn}%
  \BibitemOpen
  \bibfield  {author} {\bibinfo {author} {\bibfnamefont {L.}~\bibnamefont
  {Kofman}}, \bibinfo {author} {\bibfnamefont {A.~D.}\ \bibnamefont {Linde}}, \
  and\ \bibinfo {author} {\bibfnamefont {A.~A.}\ \bibnamefont {Starobinsky}},\
  }\href {\doibase 10.1103/PhysRevD.56.3258} {\bibfield  {journal} {\bibinfo
  {journal} {Phys. Rev. D}\ }\textbf {\bibinfo {volume} {56}},\ \bibinfo
  {pages} {3258} (\bibinfo {year} {1997})},\ \Eprint
  {http://arxiv.org/abs/hep-ph/9704452} {arXiv:hep-ph/9704452} \BibitemShut
  {NoStop}%
\bibitem [{\citenamefont {Berges}\ and\ \citenamefont
  {Serreau}(2003)}]{Berges:2002cz}%
  \BibitemOpen
  \bibfield  {author} {\bibinfo {author} {\bibfnamefont {J.}~\bibnamefont
  {Berges}}\ and\ \bibinfo {author} {\bibfnamefont {J.}~\bibnamefont
  {Serreau}},\ }\href {\doibase 10.1103/PhysRevLett.91.111601} {\bibfield
  {journal} {\bibinfo  {journal} {Phys. Rev. Lett.}\ }\textbf {\bibinfo
  {volume} {91}},\ \bibinfo {pages} {111601} (\bibinfo {year} {2003})},\
  \Eprint {http://arxiv.org/abs/hep-ph/0208070} {arXiv:hep-ph/0208070}
  \BibitemShut {NoStop}%
\bibitem [{\citenamefont {Mukhanov}(2005)}]{Mukhanov:2005sc}%
  \BibitemOpen
  \bibfield  {author} {\bibinfo {author} {\bibfnamefont {V.}~\bibnamefont
  {Mukhanov}},\ }\href@noop {} {\emph {\bibinfo {title} {{Physical Foundations
  of Cosmology}}}}\ (\bibinfo  {publisher} {Cambridge University Press},\
  \bibinfo {address} {Oxford},\ \bibinfo {year} {2005})\BibitemShut {NoStop}%
\bibitem [{\citenamefont {Dufaux}\ \emph {et~al.}(2006)\citenamefont {Dufaux},
  \citenamefont {Felder}, \citenamefont {Kofman}, \citenamefont {Peloso},\ and\
  \citenamefont {Podolsky}}]{Dufaux:2006ee}%
  \BibitemOpen
  \bibfield  {author} {\bibinfo {author} {\bibfnamefont {J.~F.}\ \bibnamefont
  {Dufaux}}, \bibinfo {author} {\bibfnamefont {G.~N.}\ \bibnamefont {Felder}},
  \bibinfo {author} {\bibfnamefont {L.}~\bibnamefont {Kofman}}, \bibinfo
  {author} {\bibfnamefont {M.}~\bibnamefont {Peloso}}, \ and\ \bibinfo {author}
  {\bibfnamefont {D.}~\bibnamefont {Podolsky}},\ }\href {\doibase
  10.1088/1475-7516/2006/07/006} {\bibfield  {journal} {\bibinfo  {journal}
  {JCAP}\ }\textbf {\bibinfo {volume} {07}},\ \bibinfo {pages} {006} (\bibinfo
  {year} {2006})},\ \Eprint {http://arxiv.org/abs/hep-ph/0602144}
  {arXiv:hep-ph/0602144} \BibitemShut {NoStop}%
\bibitem [{\citenamefont {Matsumoto}\ and\ \citenamefont
  {Moroi}(2008)}]{Matsumoto:2007rd}%
  \BibitemOpen
  \bibfield  {author} {\bibinfo {author} {\bibfnamefont {S.}~\bibnamefont
  {Matsumoto}}\ and\ \bibinfo {author} {\bibfnamefont {T.}~\bibnamefont
  {Moroi}},\ }\href {\doibase 10.1103/PhysRevD.77.045014} {\bibfield  {journal}
  {\bibinfo  {journal} {Phys. Rev. D}\ }\textbf {\bibinfo {volume} {77}},\
  \bibinfo {pages} {045014} (\bibinfo {year} {2008})},\ \Eprint
  {http://arxiv.org/abs/0709.4338} {arXiv:0709.4338 [hep-ph]} \BibitemShut
  {NoStop}%
\bibitem [{\citenamefont {Asaka}\ and\ \citenamefont
  {Nagao}(2010)}]{Asaka:2010kv}%
  \BibitemOpen
  \bibfield  {author} {\bibinfo {author} {\bibfnamefont {T.}~\bibnamefont
  {Asaka}}\ and\ \bibinfo {author} {\bibfnamefont {H.}~\bibnamefont {Nagao}},\
  }\href {\doibase 10.1143/PTP.124.293} {\bibfield  {journal} {\bibinfo
  {journal} {Prog. Theor. Phys.}\ }\textbf {\bibinfo {volume} {124}},\ \bibinfo
  {pages} {293} (\bibinfo {year} {2010})},\ \Eprint
  {http://arxiv.org/abs/1004.2125} {arXiv:1004.2125 [hep-ph]} \BibitemShut
  {NoStop}%
\bibitem [{\citenamefont {Mukaida}\ \emph {et~al.}(2013)\citenamefont
  {Mukaida}, \citenamefont {Nakayama},\ and\ \citenamefont
  {Takimoto}}]{Mukaida:2013xxa}%
  \BibitemOpen
  \bibfield  {author} {\bibinfo {author} {\bibfnamefont {K.}~\bibnamefont
  {Mukaida}}, \bibinfo {author} {\bibfnamefont {K.}~\bibnamefont {Nakayama}}, \
  and\ \bibinfo {author} {\bibfnamefont {M.}~\bibnamefont {Takimoto}},\ }\href
  {\doibase 10.1007/JHEP12(2013)053} {\bibfield  {journal} {\bibinfo  {journal}
  {JHEP}\ }\textbf {\bibinfo {volume} {12}},\ \bibinfo {pages} {053} (\bibinfo
  {year} {2013})},\ \Eprint {http://arxiv.org/abs/1308.4394} {arXiv:1308.4394
  [hep-ph]} \BibitemShut {NoStop}%
\bibitem [{\citenamefont {Amin}\ \emph
  {et~al.}(2019{\natexlab{b}})\citenamefont {Amin}, \citenamefont {Fan},
  \citenamefont {Lozanov},\ and\ \citenamefont {Reece}}]{Amin:2019qrx}%
  \BibitemOpen
  \bibfield  {author} {\bibinfo {author} {\bibfnamefont {M.~A.}\ \bibnamefont
  {Amin}}, \bibinfo {author} {\bibfnamefont {J.}~\bibnamefont {Fan}}, \bibinfo
  {author} {\bibfnamefont {K.~D.}\ \bibnamefont {Lozanov}}, \ and\ \bibinfo
  {author} {\bibfnamefont {M.}~\bibnamefont {Reece}},\ }\href {\doibase
  10.1103/PhysRevD.99.035008} {\bibfield  {journal} {\bibinfo  {journal} {Phys.
  Rev. D}\ }\textbf {\bibinfo {volume} {99}},\ \bibinfo {pages} {035008}
  (\bibinfo {year} {2019}{\natexlab{b}})},\ \Eprint
  {http://arxiv.org/abs/1802.00444} {arXiv:1802.00444 [hep-ph]} \BibitemShut
  {NoStop}%
\bibitem [{\citenamefont {Kitajima}\ \emph
  {et~al.}(2018{\natexlab{b}})\citenamefont {Kitajima}, \citenamefont
  {Sekiguchi},\ and\ \citenamefont {Takahashi}}]{Kitajima:2017peg}%
  \BibitemOpen
  \bibfield  {author} {\bibinfo {author} {\bibfnamefont {N.}~\bibnamefont
  {Kitajima}}, \bibinfo {author} {\bibfnamefont {T.}~\bibnamefont {Sekiguchi}},
  \ and\ \bibinfo {author} {\bibfnamefont {F.}~\bibnamefont {Takahashi}},\
  }\href {\doibase 10.1016/j.physletb.2018.04.024} {\bibfield  {journal}
  {\bibinfo  {journal} {Phys. Lett. B}\ }\textbf {\bibinfo {volume} {781}},\
  \bibinfo {pages} {684} (\bibinfo {year} {2018}{\natexlab{b}})},\ \Eprint
  {http://arxiv.org/abs/1711.06590} {arXiv:1711.06590 [hep-ph]} \BibitemShut
  {NoStop}%
\bibitem [{\citenamefont {Agrawal}\ \emph {et~al.}(2020)\citenamefont
  {Agrawal}, \citenamefont {Kitajima}, \citenamefont {Reece}, \citenamefont
  {Sekiguchi},\ and\ \citenamefont {Takahashi}}]{Agrawal:2018vin}%
  \BibitemOpen
  \bibfield  {author} {\bibinfo {author} {\bibfnamefont {P.}~\bibnamefont
  {Agrawal}}, \bibinfo {author} {\bibfnamefont {N.}~\bibnamefont {Kitajima}},
  \bibinfo {author} {\bibfnamefont {M.}~\bibnamefont {Reece}}, \bibinfo
  {author} {\bibfnamefont {T.}~\bibnamefont {Sekiguchi}}, \ and\ \bibinfo
  {author} {\bibfnamefont {F.}~\bibnamefont {Takahashi}},\ }\href {\doibase
  10.1016/j.physletb.2019.135136} {\bibfield  {journal} {\bibinfo  {journal}
  {Phys. Lett. B}\ }\textbf {\bibinfo {volume} {801}},\ \bibinfo {pages}
  {135136} (\bibinfo {year} {2020})},\ \Eprint
  {http://arxiv.org/abs/1810.07188} {arXiv:1810.07188 [hep-ph]} \BibitemShut
  {NoStop}%
\bibitem [{\citenamefont {Co}\ \emph {et~al.}(2019)\citenamefont {Co},
  \citenamefont {Pierce}, \citenamefont {Zhang},\ and\ \citenamefont
  {Zhao}}]{Co:2018lka}%
  \BibitemOpen
  \bibfield  {author} {\bibinfo {author} {\bibfnamefont {R.~T.}\ \bibnamefont
  {Co}}, \bibinfo {author} {\bibfnamefont {A.}~\bibnamefont {Pierce}}, \bibinfo
  {author} {\bibfnamefont {Z.}~\bibnamefont {Zhang}}, \ and\ \bibinfo {author}
  {\bibfnamefont {Y.}~\bibnamefont {Zhao}},\ }\href {\doibase
  10.1103/PhysRevD.99.075002} {\bibfield  {journal} {\bibinfo  {journal} {Phys.
  Rev. D}\ }\textbf {\bibinfo {volume} {99}},\ \bibinfo {pages} {075002}
  (\bibinfo {year} {2019})},\ \Eprint {http://arxiv.org/abs/1810.07196}
  {arXiv:1810.07196 [hep-ph]} \BibitemShut {NoStop}%
\bibitem [{\citenamefont {Dror}\ \emph {et~al.}(2019)\citenamefont {Dror},
  \citenamefont {Harigaya},\ and\ \citenamefont {Narayan}}]{Dror:2018pdh}%
  \BibitemOpen
  \bibfield  {author} {\bibinfo {author} {\bibfnamefont {J.~A.}\ \bibnamefont
  {Dror}}, \bibinfo {author} {\bibfnamefont {K.}~\bibnamefont {Harigaya}}, \
  and\ \bibinfo {author} {\bibfnamefont {V.}~\bibnamefont {Narayan}},\ }\href
  {\doibase 10.1103/PhysRevD.99.035036} {\bibfield  {journal} {\bibinfo
  {journal} {Phys. Rev. D}\ }\textbf {\bibinfo {volume} {99}},\ \bibinfo
  {pages} {035036} (\bibinfo {year} {2019})},\ \Eprint
  {http://arxiv.org/abs/1810.07195} {arXiv:1810.07195 [hep-ph]} \BibitemShut
  {NoStop}%
\bibitem [{\citenamefont {Lozanov}(2019)}]{Lozanov:2019jxc}%
  \BibitemOpen
  \bibfield  {author} {\bibinfo {author} {\bibfnamefont {K.~D.}\ \bibnamefont
  {Lozanov}},\ }\href@noop {} {\  (\bibinfo {year} {2019})},\ \Eprint
  {http://arxiv.org/abs/1907.04402} {arXiv:1907.04402 [astro-ph.CO]}
  \BibitemShut {NoStop}%
\bibitem [{\citenamefont {Alonso-\'Alvarez}\ \emph {et~al.}(2020)\citenamefont
  {Alonso-\'Alvarez}, \citenamefont {Gupta}, \citenamefont {Jaeckel},\ and\
  \citenamefont {Spannowsky}}]{Alonso-Alvarez:2019ssa}%
  \BibitemOpen
  \bibfield  {author} {\bibinfo {author} {\bibfnamefont {G.}~\bibnamefont
  {Alonso-\'Alvarez}}, \bibinfo {author} {\bibfnamefont {R.~S.}\ \bibnamefont
  {Gupta}}, \bibinfo {author} {\bibfnamefont {J.}~\bibnamefont {Jaeckel}}, \
  and\ \bibinfo {author} {\bibfnamefont {M.}~\bibnamefont {Spannowsky}},\
  }\href {\doibase 10.1088/1475-7516/2020/03/052} {\bibfield  {journal}
  {\bibinfo  {journal} {JCAP}\ }\textbf {\bibinfo {volume} {03}},\ \bibinfo
  {pages} {052} (\bibinfo {year} {2020})},\ \Eprint
  {http://arxiv.org/abs/1911.07885} {arXiv:1911.07885 [hep-ph]} \BibitemShut
  {NoStop}%
\bibitem [{\citenamefont {Moroi}\ and\ \citenamefont
  {Yin}(2020{\natexlab{a}})}]{Moroi:2020bkq}%
  \BibitemOpen
  \bibfield  {author} {\bibinfo {author} {\bibfnamefont {T.}~\bibnamefont
  {Moroi}}\ and\ \bibinfo {author} {\bibfnamefont {W.}~\bibnamefont {Yin}},\
  }\href@noop {} {\  (\bibinfo {year} {2020}{\natexlab{a}})},\ \Eprint
  {http://arxiv.org/abs/2011.12285} {arXiv:2011.12285 [hep-ph]} \BibitemShut
  {NoStop}%
\bibitem [{\citenamefont {Irastorza}\ \emph {et~al.}(2011)\citenamefont
  {Irastorza} \emph {et~al.}}]{Irastorza:2011gs}%
  \BibitemOpen
  \bibfield  {author} {\bibinfo {author} {\bibfnamefont {I.}~\bibnamefont
  {Irastorza}} \emph {et~al.},\ }\href {\doibase 10.1088/1475-7516/2011/06/013}
  {\bibfield  {journal} {\bibinfo  {journal} {JCAP}\ }\textbf {\bibinfo
  {volume} {06}},\ \bibinfo {pages} {013} (\bibinfo {year} {2011})},\ \Eprint
  {http://arxiv.org/abs/1103.5334} {arXiv:1103.5334 [hep-ex]} \BibitemShut
  {NoStop}%
\bibitem [{\citenamefont {Armengaud}\ \emph {et~al.}(2014)\citenamefont
  {Armengaud} \emph {et~al.}}]{Armengaud:2014gea}%
  \BibitemOpen
  \bibfield  {author} {\bibinfo {author} {\bibfnamefont {E.}~\bibnamefont
  {Armengaud}} \emph {et~al.},\ }\href {\doibase 10.1088/1748-0221/9/05/T05002}
  {\bibfield  {journal} {\bibinfo  {journal} {JINST}\ }\textbf {\bibinfo
  {volume} {9}},\ \bibinfo {pages} {T05002} (\bibinfo {year} {2014})},\ \Eprint
  {http://arxiv.org/abs/1401.3233} {arXiv:1401.3233 [physics.ins-det]}
  \BibitemShut {NoStop}%
\bibitem [{\citenamefont {Abeln}\ \emph {et~al.}(2020)\citenamefont {Abeln}
  \emph {et~al.}}]{Abeln:2020ywv}%
  \BibitemOpen
  \bibfield  {author} {\bibinfo {author} {\bibfnamefont {A.}~\bibnamefont
  {Abeln}} \emph {et~al.} (\bibinfo {collaboration} {BabyIAXO}),\ }\href@noop
  {} {\  (\bibinfo {year} {2020})},\ \Eprint {http://arxiv.org/abs/2010.12076}
  {arXiv:2010.12076 [physics.ins-det]} \BibitemShut {NoStop}%
\bibitem [{\citenamefont {{Wiebusch}}(2009)}]{2009arXiv0907.2263W}%
  \BibitemOpen
  \bibfield  {author} {\bibinfo {author} {\bibfnamefont {C.}~\bibnamefont
  {{Wiebusch}}},\ }\href@noop {} {\bibfield  {journal} {\bibinfo  {journal}
  {arXiv e-prints}\ ,\ \bibinfo {eid} {arXiv:0907.2263}} (\bibinfo {year}
  {2009})},\ \Eprint {http://arxiv.org/abs/0907.2263} {arXiv:0907.2263
  [astro-ph.IM]} \BibitemShut {NoStop}%
\bibitem [{\citenamefont {Aartsen}\ \emph
  {et~al.}(2014{\natexlab{a}})\citenamefont {Aartsen} \emph
  {et~al.}}]{Aartsen:2014gkd}%
  \BibitemOpen
  \bibfield  {author} {\bibinfo {author} {\bibfnamefont {M.}~\bibnamefont
  {Aartsen}} \emph {et~al.} (\bibinfo {collaboration} {IceCube}),\ }\href
  {\doibase 10.1103/PhysRevLett.113.101101} {\bibfield  {journal} {\bibinfo
  {journal} {Phys. Rev. Lett.}\ }\textbf {\bibinfo {volume} {113}},\ \bibinfo
  {pages} {101101} (\bibinfo {year} {2014}{\natexlab{a}})},\ \Eprint
  {http://arxiv.org/abs/1405.5303} {arXiv:1405.5303 [astro-ph.HE]} \BibitemShut
  {NoStop}%
\bibitem [{\citenamefont {Aartsen}\ \emph
  {et~al.}(2014{\natexlab{b}})\citenamefont {Aartsen} \emph
  {et~al.}}]{Aartsen:2014njl}%
  \BibitemOpen
  \bibfield  {author} {\bibinfo {author} {\bibfnamefont {M.~G.}\ \bibnamefont
  {Aartsen}} \emph {et~al.} (\bibinfo {collaboration} {IceCube}),\ }\href@noop
  {} {\  (\bibinfo {year} {2014}{\natexlab{b}})},\ \Eprint
  {http://arxiv.org/abs/1412.5106} {arXiv:1412.5106 [astro-ph.HE]} \BibitemShut
  {NoStop}%
\bibitem [{\citenamefont {Aartsen}\ \emph {et~al.}(2020)\citenamefont {Aartsen}
  \emph {et~al.}}]{Aartsen:2020aqd}%
  \BibitemOpen
  \bibfield  {author} {\bibinfo {author} {\bibfnamefont {M.}~\bibnamefont
  {Aartsen}} \emph {et~al.} (\bibinfo {collaboration} {IceCube}),\ }\href
  {\doibase 10.1103/PhysRevLett.125.121104} {\bibfield  {journal} {\bibinfo
  {journal} {Phys. Rev. Lett.}\ }\textbf {\bibinfo {volume} {125}},\ \bibinfo
  {pages} {121104} (\bibinfo {year} {2020})},\ \Eprint
  {http://arxiv.org/abs/2001.09520} {arXiv:2001.09520 [astro-ph.HE]}
  \BibitemShut {NoStop}%
\bibitem [{\citenamefont {Abbasi}\ \emph {et~al.}(2020)\citenamefont {Abbasi}
  \emph {et~al.}}]{Abbasi:2020jmh}%
  \BibitemOpen
  \bibfield  {author} {\bibinfo {author} {\bibfnamefont {R.}~\bibnamefont
  {Abbasi}} \emph {et~al.} (\bibinfo {collaboration} {IceCube}),\ }\href@noop
  {} {\  (\bibinfo {year} {2020})},\ \Eprint {http://arxiv.org/abs/2011.03545}
  {arXiv:2011.03545 [astro-ph.HE]} \BibitemShut {NoStop}%
\bibitem [{\citenamefont {Aalbers}\ \emph {et~al.}(2016)\citenamefont {Aalbers}
  \emph {et~al.}}]{Aalbers:2016jon}%
  \BibitemOpen
  \bibfield  {author} {\bibinfo {author} {\bibfnamefont {J.}~\bibnamefont
  {Aalbers}} \emph {et~al.} (\bibinfo {collaboration} {DARWIN}),\ }\href
  {\doibase 10.1088/1475-7516/2016/11/017} {\bibfield  {journal} {\bibinfo
  {journal} {JCAP}\ }\textbf {\bibinfo {volume} {11}},\ \bibinfo {pages} {017}
  (\bibinfo {year} {2016})},\ \Eprint {http://arxiv.org/abs/1606.07001}
  {arXiv:1606.07001 [astro-ph.IM]} \BibitemShut {NoStop}%
\bibitem [{\citenamefont {Lazarides}\ and\ \citenamefont
  {Shafi}(1991)}]{Lazarides:1991wu}%
  \BibitemOpen
  \bibfield  {author} {\bibinfo {author} {\bibfnamefont {G.}~\bibnamefont
  {Lazarides}}\ and\ \bibinfo {author} {\bibfnamefont {Q.}~\bibnamefont
  {Shafi}},\ }\href {\doibase 10.1016/0370-2693(91)91090-I} {\bibfield
  {journal} {\bibinfo  {journal} {Phys. Lett. B}\ }\textbf {\bibinfo {volume}
  {258}},\ \bibinfo {pages} {305} (\bibinfo {year} {1991})}\BibitemShut
  {NoStop}%
\bibitem [{\citenamefont {Nakayama}\ and\ \citenamefont
  {Takahashi}(2011)}]{Nakayama:2011ri}%
  \BibitemOpen
  \bibfield  {author} {\bibinfo {author} {\bibfnamefont {K.}~\bibnamefont
  {Nakayama}}\ and\ \bibinfo {author} {\bibfnamefont {F.}~\bibnamefont
  {Takahashi}},\ }\href {\doibase 10.1088/1475-7516/2011/10/033} {\bibfield
  {journal} {\bibinfo  {journal} {JCAP}\ }\textbf {\bibinfo {volume} {10}},\
  \bibinfo {pages} {033} (\bibinfo {year} {2011})},\ \Eprint
  {http://arxiv.org/abs/1108.0070} {arXiv:1108.0070 [hep-ph]} \BibitemShut
  {NoStop}%
\bibitem [{\citenamefont {King}\ and\ \citenamefont
  {Ludl}(2017)}]{King:2017nbl}%
  \BibitemOpen
  \bibfield  {author} {\bibinfo {author} {\bibfnamefont {S.~F.}\ \bibnamefont
  {King}}\ and\ \bibinfo {author} {\bibfnamefont {P.~O.}\ \bibnamefont
  {Ludl}},\ }\href {\doibase 10.1007/JHEP03(2017)174} {\bibfield  {journal}
  {\bibinfo  {journal} {JHEP}\ }\textbf {\bibinfo {volume} {03}},\ \bibinfo
  {pages} {174} (\bibinfo {year} {2017})},\ \Eprint
  {http://arxiv.org/abs/1701.04794} {arXiv:1701.04794 [hep-ph]} \BibitemShut
  {NoStop}%
\bibitem [{\citenamefont {Antusch}\ and\ \citenamefont
  {Marschall}(2018)}]{Antusch:2018zvu}%
  \BibitemOpen
  \bibfield  {author} {\bibinfo {author} {\bibfnamefont {S.}~\bibnamefont
  {Antusch}}\ and\ \bibinfo {author} {\bibfnamefont {K.}~\bibnamefont
  {Marschall}},\ }\href {\doibase 10.1088/1475-7516/2018/05/015} {\bibfield
  {journal} {\bibinfo  {journal} {JCAP}\ }\textbf {\bibinfo {volume} {05}},\
  \bibinfo {pages} {015} (\bibinfo {year} {2018})},\ \Eprint
  {http://arxiv.org/abs/1802.05647} {arXiv:1802.05647 [hep-ph]} \BibitemShut
  {NoStop}%
\bibitem [{\citenamefont {Takahashi}\ \emph {et~al.}(2018)\citenamefont
  {Takahashi}, \citenamefont {Yin},\ and\ \citenamefont {Guth}}]{Guth:2018hsa}%
  \BibitemOpen
  \bibfield  {author} {\bibinfo {author} {\bibfnamefont {F.}~\bibnamefont
  {Takahashi}}, \bibinfo {author} {\bibfnamefont {W.}~\bibnamefont {Yin}}, \
  and\ \bibinfo {author} {\bibfnamefont {A.~H.}\ \bibnamefont {Guth}},\ }\href
  {\doibase 10.1103/PhysRevD.98.015042} {\bibfield  {journal} {\bibinfo
  {journal} {Phys. Rev.}\ }\textbf {\bibinfo {volume} {D98}},\ \bibinfo {pages}
  {015042} (\bibinfo {year} {2018})},\ \Eprint
  {http://arxiv.org/abs/1805.08763} {arXiv:1805.08763 [hep-ph]} \BibitemShut
  {NoStop}%
%%CITATION = ARXIV:1805.08763;%%
\bibitem [{\citenamefont {Aghanim}\ \emph {et~al.}(2018)\citenamefont {Aghanim}
  \emph {et~al.}}]{Aghanim:2018eyx}%
  \BibitemOpen
  \bibfield  {author} {\bibinfo {author} {\bibfnamefont {N.}~\bibnamefont
  {Aghanim}} \emph {et~al.} (\bibinfo {collaboration} {Planck}),\ }\href@noop
  {} {\  (\bibinfo {year} {2018})},\ \Eprint {http://arxiv.org/abs/1807.06209}
  {arXiv:1807.06209 [astro-ph.CO]} \BibitemShut {NoStop}%
%%CITATION = ARXIV:1807.06209;%%
\bibitem [{\citenamefont {Fields}\ \emph {et~al.}(2020)\citenamefont {Fields},
  \citenamefont {Olive}, \citenamefont {Yeh},\ and\ \citenamefont
  {Young}}]{Fields:2019pfx}%
  \BibitemOpen
  \bibfield  {author} {\bibinfo {author} {\bibfnamefont {B.~D.}\ \bibnamefont
  {Fields}}, \bibinfo {author} {\bibfnamefont {K.~A.}\ \bibnamefont {Olive}},
  \bibinfo {author} {\bibfnamefont {T.-H.}\ \bibnamefont {Yeh}}, \ and\
  \bibinfo {author} {\bibfnamefont {C.}~\bibnamefont {Young}},\ }\href
  {\doibase 10.1088/1475-7516/2020/03/010} {\bibfield  {journal} {\bibinfo
  {journal} {JCAP}\ }\textbf {\bibinfo {volume} {03}},\ \bibinfo {pages} {010}
  (\bibinfo {year} {2020})},\ \bibinfo {note} {[Erratum: JCAP 11, E02
  (2020)]},\ \Eprint {http://arxiv.org/abs/1912.01132} {arXiv:1912.01132
  [astro-ph.CO]} \BibitemShut {NoStop}%
\bibitem [{\citenamefont {Husdal}(2016)}]{Husdal:2016haj}%
  \BibitemOpen
  \bibfield  {author} {\bibinfo {author} {\bibfnamefont {L.}~\bibnamefont
  {Husdal}},\ }\href {\doibase 10.3390/galaxies4040078} {\bibfield  {journal}
  {\bibinfo  {journal} {Galaxies}\ }\textbf {\bibinfo {volume} {4}},\ \bibinfo
  {pages} {78} (\bibinfo {year} {2016})},\ \Eprint
  {http://arxiv.org/abs/1609.04979} {arXiv:1609.04979 [astro-ph.CO]}
  \BibitemShut {NoStop}%
%%CITATION = ARXIV:1609.04979;%%
\bibitem [{\citenamefont {Moroi}\ and\ \citenamefont
  {Yin}(2020{\natexlab{b}})}]{Moroi:2020has}%
  \BibitemOpen
  \bibfield  {author} {\bibinfo {author} {\bibfnamefont {T.}~\bibnamefont
  {Moroi}}\ and\ \bibinfo {author} {\bibfnamefont {W.}~\bibnamefont {Yin}},\
  }\href@noop {} {\  (\bibinfo {year} {2020}{\natexlab{b}})},\ \Eprint
  {http://arxiv.org/abs/2011.09475} {arXiv:2011.09475 [hep-ph]} \BibitemShut
  {NoStop}%
\bibitem [{\citenamefont {Arguelles}\ \emph {et~al.}(2019)\citenamefont
  {Arguelles} \emph {et~al.}}]{Arguelles:2019xgp}%
  \BibitemOpen
  \bibfield  {author} {\bibinfo {author} {\bibfnamefont {C.~A.}\ \bibnamefont
  {Arguelles}} \emph {et~al.},\ }\href@noop {} {\  (\bibinfo {year} {2019})},\
  \Eprint {http://arxiv.org/abs/1907.08311} {arXiv:1907.08311 [hep-ph]}
  \BibitemShut {NoStop}%
%%CITATION = ARXIV:1907.08311;%%
\bibitem [{\citenamefont {Aartsen}\ \emph {et~al.}(2017)\citenamefont {Aartsen}
  \emph {et~al.}}]{Aartsen:2017mau}%
  \BibitemOpen
  \bibfield  {author} {\bibinfo {author} {\bibfnamefont {M.~G.}\ \bibnamefont
  {Aartsen}} \emph {et~al.} (\bibinfo {collaboration} {IceCube}),\ }\href@noop
  {} {\  (\bibinfo {year} {2017})},\ \Eprint {http://arxiv.org/abs/1710.01191}
  {arXiv:1710.01191 [astro-ph.HE]} \BibitemShut {NoStop}%
%%CITATION = ARXIV:1710.01191;%%
\bibitem [{\citenamefont {Aartsen}\ \emph {et~al.}(2018)\citenamefont {Aartsen}
  \emph {et~al.}}]{Aartsen:2018vtx}%
  \BibitemOpen
  \bibfield  {author} {\bibinfo {author} {\bibfnamefont {M.~G.}\ \bibnamefont
  {Aartsen}} \emph {et~al.} (\bibinfo {collaboration} {IceCube}),\ }\href
  {\doibase 10.1103/PhysRevD.98.062003} {\bibfield  {journal} {\bibinfo
  {journal} {Phys. Rev.}\ }\textbf {\bibinfo {volume} {D98}},\ \bibinfo {pages}
  {062003} (\bibinfo {year} {2018})},\ \Eprint
  {http://arxiv.org/abs/1807.01820} {arXiv:1807.01820 [astro-ph.HE]}
  \BibitemShut {NoStop}%
%%CITATION = ARXIV:1807.01820;%%
\bibitem [{\citenamefont {Gorham}\ \emph {et~al.}(2016)\citenamefont {Gorham}
  \emph {et~al.}}]{Gorham:2016zah}%
  \BibitemOpen
  \bibfield  {author} {\bibinfo {author} {\bibfnamefont {P.~W.}\ \bibnamefont
  {Gorham}} \emph {et~al.} (\bibinfo {collaboration} {ANITA}),\ }\href
  {\doibase 10.1103/PhysRevLett.117.071101} {\bibfield  {journal} {\bibinfo
  {journal} {Phys. Rev. Lett.}\ }\textbf {\bibinfo {volume} {117}},\ \bibinfo
  {pages} {071101} (\bibinfo {year} {2016})},\ \Eprint
  {http://arxiv.org/abs/1603.05218} {arXiv:1603.05218 [astro-ph.HE]}
  \BibitemShut {NoStop}%
%%CITATION = ARXIV:1603.05218;%%
\bibitem [{\citenamefont {Gorham}\ \emph {et~al.}(2018)\citenamefont {Gorham}
  \emph {et~al.}}]{Gorham:2018ydl}%
  \BibitemOpen
  \bibfield  {author} {\bibinfo {author} {\bibfnamefont {P.~W.}\ \bibnamefont
  {Gorham}} \emph {et~al.} (\bibinfo {collaboration} {ANITA}),\ }\href
  {\doibase 10.1103/PhysRevLett.121.161102} {\bibfield  {journal} {\bibinfo
  {journal} {Phys. Rev. Lett.}\ }\textbf {\bibinfo {volume} {121}},\ \bibinfo
  {pages} {161102} (\bibinfo {year} {2018})},\ \Eprint
  {http://arxiv.org/abs/1803.05088} {arXiv:1803.05088 [astro-ph.HE]}
  \BibitemShut {NoStop}%
%%CITATION = ARXIV:1803.05088;%%
\bibitem [{\citenamefont {Aprile}\ \emph {et~al.}(2020)\citenamefont {Aprile}
  \emph {et~al.}}]{Aprile:2020tmw}%
  \BibitemOpen
  \bibfield  {author} {\bibinfo {author} {\bibfnamefont {E.}~\bibnamefont
  {Aprile}} \emph {et~al.} (\bibinfo {collaboration} {XENON}),\ }\href@noop {}
  {\  (\bibinfo {year} {2020})},\ \Eprint {http://arxiv.org/abs/2006.09721}
  {arXiv:2006.09721 [hep-ex]} \BibitemShut {NoStop}%
\bibitem [{\citenamefont {Feldstein}\ \emph {et~al.}(2013)\citenamefont
  {Feldstein}, \citenamefont {Kusenko}, \citenamefont {Matsumoto},\ and\
  \citenamefont {Yanagida}}]{Feldstein:2013kka}%
  \BibitemOpen
  \bibfield  {author} {\bibinfo {author} {\bibfnamefont {B.}~\bibnamefont
  {Feldstein}}, \bibinfo {author} {\bibfnamefont {A.}~\bibnamefont {Kusenko}},
  \bibinfo {author} {\bibfnamefont {S.}~\bibnamefont {Matsumoto}}, \ and\
  \bibinfo {author} {\bibfnamefont {T.~T.}\ \bibnamefont {Yanagida}},\ }\href
  {\doibase 10.1103/PhysRevD.88.015004} {\bibfield  {journal} {\bibinfo
  {journal} {Phys. Rev.}\ }\textbf {\bibinfo {volume} {D88}},\ \bibinfo {pages}
  {015004} (\bibinfo {year} {2013})},\ \Eprint {http://arxiv.org/abs/1303.7320}
  {arXiv:1303.7320 [hep-ph]} \BibitemShut {NoStop}%
%%CITATION = ARXIV:1303.7320;%%
\bibitem [{\citenamefont {Esmaili}\ and\ \citenamefont
  {Serpico}(2013)}]{Esmaili:2013gha}%
  \BibitemOpen
  \bibfield  {author} {\bibinfo {author} {\bibfnamefont {A.}~\bibnamefont
  {Esmaili}}\ and\ \bibinfo {author} {\bibfnamefont {P.~D.}\ \bibnamefont
  {Serpico}},\ }\href {\doibase 10.1088/1475-7516/2013/11/054} {\bibfield
  {journal} {\bibinfo  {journal} {JCAP}\ }\textbf {\bibinfo {volume} {1311}},\
  \bibinfo {pages} {054} (\bibinfo {year} {2013})},\ \Eprint
  {http://arxiv.org/abs/1308.1105} {arXiv:1308.1105 [hep-ph]} \BibitemShut
  {NoStop}%
%%CITATION = ARXIV:1308.1105;%%
\bibitem [{\citenamefont {Higaki}\ \emph {et~al.}(2014)\citenamefont {Higaki},
  \citenamefont {Kitano},\ and\ \citenamefont {Sato}}]{Higaki:2014dwa}%
  \BibitemOpen
  \bibfield  {author} {\bibinfo {author} {\bibfnamefont {T.}~\bibnamefont
  {Higaki}}, \bibinfo {author} {\bibfnamefont {R.}~\bibnamefont {Kitano}}, \
  and\ \bibinfo {author} {\bibfnamefont {R.}~\bibnamefont {Sato}},\ }\href
  {\doibase 10.1007/JHEP07(2014)044} {\bibfield  {journal} {\bibinfo  {journal}
  {JHEP}\ }\textbf {\bibinfo {volume} {07}},\ \bibinfo {pages} {044} (\bibinfo
  {year} {2014})},\ \Eprint {http://arxiv.org/abs/1405.0013} {arXiv:1405.0013
  [hep-ph]} \BibitemShut {NoStop}%
%%CITATION = ARXIV:1405.0013;%%
\bibitem [{\citenamefont {Rott}\ \emph {et~al.}(2015)\citenamefont {Rott},
  \citenamefont {Kohri},\ and\ \citenamefont {Park}}]{Rott:2014kfa}%
  \BibitemOpen
  \bibfield  {author} {\bibinfo {author} {\bibfnamefont {C.}~\bibnamefont
  {Rott}}, \bibinfo {author} {\bibfnamefont {K.}~\bibnamefont {Kohri}}, \ and\
  \bibinfo {author} {\bibfnamefont {S.~C.}\ \bibnamefont {Park}},\ }\href
  {\doibase 10.1103/PhysRevD.92.023529} {\bibfield  {journal} {\bibinfo
  {journal} {Phys. Rev.}\ }\textbf {\bibinfo {volume} {D92}},\ \bibinfo {pages}
  {023529} (\bibinfo {year} {2015})},\ \Eprint {http://arxiv.org/abs/1408.4575}
  {arXiv:1408.4575 [hep-ph]} \BibitemShut {NoStop}%
%%CITATION = ARXIV:1408.4575;%%
\bibitem [{\citenamefont {Dudas}\ \emph {et~al.}(2015)\citenamefont {Dudas},
  \citenamefont {Mambrini},\ and\ \citenamefont {Olive}}]{Dudas:2014bca}%
  \BibitemOpen
  \bibfield  {author} {\bibinfo {author} {\bibfnamefont {E.}~\bibnamefont
  {Dudas}}, \bibinfo {author} {\bibfnamefont {Y.}~\bibnamefont {Mambrini}}, \
  and\ \bibinfo {author} {\bibfnamefont {K.~A.}\ \bibnamefont {Olive}},\ }\href
  {\doibase 10.1103/PhysRevD.91.075001} {\bibfield  {journal} {\bibinfo
  {journal} {Phys. Rev.}\ }\textbf {\bibinfo {volume} {D91}},\ \bibinfo {pages}
  {075001} (\bibinfo {year} {2015})},\ \Eprint {http://arxiv.org/abs/1412.3459}
  {arXiv:1412.3459 [hep-ph]} \BibitemShut {NoStop}%
%%CITATION = ARXIV:1412.3459;%%
\bibitem [{\citenamefont {Murase}\ \emph {et~al.}(2015)\citenamefont {Murase},
  \citenamefont {Laha}, \citenamefont {Ando},\ and\ \citenamefont
  {Ahlers}}]{Murase:2015gea}%
  \BibitemOpen
  \bibfield  {author} {\bibinfo {author} {\bibfnamefont {K.}~\bibnamefont
  {Murase}}, \bibinfo {author} {\bibfnamefont {R.}~\bibnamefont {Laha}},
  \bibinfo {author} {\bibfnamefont {S.}~\bibnamefont {Ando}}, \ and\ \bibinfo
  {author} {\bibfnamefont {M.}~\bibnamefont {Ahlers}},\ }\href {\doibase
  10.1103/PhysRevLett.115.071301} {\bibfield  {journal} {\bibinfo  {journal}
  {Phys. Rev. Lett.}\ }\textbf {\bibinfo {volume} {115}},\ \bibinfo {pages}
  {071301} (\bibinfo {year} {2015})},\ \Eprint
  {http://arxiv.org/abs/1503.04663} {arXiv:1503.04663 [hep-ph]} \BibitemShut
  {NoStop}%
%%CITATION = ARXIV:1503.04663;%%
\bibitem [{\citenamefont {Dev}\ \emph {et~al.}(2016)\citenamefont {Dev},
  \citenamefont {Kazanas}, \citenamefont {Mohapatra}, \citenamefont {Teplitz},\
  and\ \citenamefont {Zhang}}]{Dev:2016qbd}%
  \BibitemOpen
  \bibfield  {author} {\bibinfo {author} {\bibfnamefont {P.~S.~B.}\
  \bibnamefont {Dev}}, \bibinfo {author} {\bibfnamefont {D.}~\bibnamefont
  {Kazanas}}, \bibinfo {author} {\bibfnamefont {R.~N.}\ \bibnamefont
  {Mohapatra}}, \bibinfo {author} {\bibfnamefont {V.~L.}\ \bibnamefont
  {Teplitz}}, \ and\ \bibinfo {author} {\bibfnamefont {Y.}~\bibnamefont
  {Zhang}},\ }\href {\doibase 10.1088/1475-7516/2016/08/034} {\bibfield
  {journal} {\bibinfo  {journal} {JCAP}\ }\textbf {\bibinfo {volume} {1608}},\
  \bibinfo {pages} {034} (\bibinfo {year} {2016})},\ \Eprint
  {http://arxiv.org/abs/1606.04517} {arXiv:1606.04517 [hep-ph]} \BibitemShut
  {NoStop}%
%%CITATION = ARXIV:1606.04517;%%
\bibitem [{\citenamefont {Hiroshima}\ \emph {et~al.}(2018)\citenamefont
  {Hiroshima}, \citenamefont {Kitano}, \citenamefont {Kohri},\ and\
  \citenamefont {Murase}}]{Hiroshima:2017hmy}%
  \BibitemOpen
  \bibfield  {author} {\bibinfo {author} {\bibfnamefont {N.}~\bibnamefont
  {Hiroshima}}, \bibinfo {author} {\bibfnamefont {R.}~\bibnamefont {Kitano}},
  \bibinfo {author} {\bibfnamefont {K.}~\bibnamefont {Kohri}}, \ and\ \bibinfo
  {author} {\bibfnamefont {K.}~\bibnamefont {Murase}},\ }\href {\doibase
  10.1103/PhysRevD.97.023006} {\bibfield  {journal} {\bibinfo  {journal} {Phys.
  Rev.}\ }\textbf {\bibinfo {volume} {D97}},\ \bibinfo {pages} {023006}
  (\bibinfo {year} {2018})},\ \Eprint {http://arxiv.org/abs/1705.04419}
  {arXiv:1705.04419 [hep-ph]} \BibitemShut {NoStop}%
%%CITATION = ARXIV:1705.04419;%%
\bibitem [{\citenamefont {Bhattacharya}\ \emph {et~al.}(2015)\citenamefont
  {Bhattacharya}, \citenamefont {Gandhi},\ and\ \citenamefont
  {Gupta}}]{Bhattacharya:2014yha}%
  \BibitemOpen
  \bibfield  {author} {\bibinfo {author} {\bibfnamefont {A.}~\bibnamefont
  {Bhattacharya}}, \bibinfo {author} {\bibfnamefont {R.}~\bibnamefont
  {Gandhi}}, \ and\ \bibinfo {author} {\bibfnamefont {A.}~\bibnamefont
  {Gupta}},\ }\href {\doibase 10.1088/1475-7516/2015/03/027} {\bibfield
  {journal} {\bibinfo  {journal} {JCAP}\ }\textbf {\bibinfo {volume} {1503}},\
  \bibinfo {pages} {027} (\bibinfo {year} {2015})},\ \Eprint
  {http://arxiv.org/abs/1407.3280} {arXiv:1407.3280 [hep-ph]} \BibitemShut
  {NoStop}%
%%CITATION = ARXIV:1407.3280;%%
\bibitem [{\citenamefont {Kopp}\ \emph {et~al.}(2015)\citenamefont {Kopp},
  \citenamefont {Liu},\ and\ \citenamefont {Wang}}]{Kopp:2015bfa}%
  \BibitemOpen
  \bibfield  {author} {\bibinfo {author} {\bibfnamefont {J.}~\bibnamefont
  {Kopp}}, \bibinfo {author} {\bibfnamefont {J.}~\bibnamefont {Liu}}, \ and\
  \bibinfo {author} {\bibfnamefont {X.-P.}\ \bibnamefont {Wang}},\ }\href
  {\doibase 10.1007/JHEP04(2015)105} {\bibfield  {journal} {\bibinfo  {journal}
  {JHEP}\ }\textbf {\bibinfo {volume} {04}},\ \bibinfo {pages} {105} (\bibinfo
  {year} {2015})},\ \Eprint {http://arxiv.org/abs/1503.02669} {arXiv:1503.02669
  [hep-ph]} \BibitemShut {NoStop}%
%%CITATION = ARXIV:1503.02669;%%
\bibitem [{\citenamefont {Cui}\ \emph {et~al.}(2018)\citenamefont {Cui},
  \citenamefont {Pospelov},\ and\ \citenamefont {Pradler}}]{Cui:2017ytb}%
  \BibitemOpen
  \bibfield  {author} {\bibinfo {author} {\bibfnamefont {Y.}~\bibnamefont
  {Cui}}, \bibinfo {author} {\bibfnamefont {M.}~\bibnamefont {Pospelov}}, \
  and\ \bibinfo {author} {\bibfnamefont {J.}~\bibnamefont {Pradler}},\ }\href
  {\doibase 10.1103/PhysRevD.97.103004} {\bibfield  {journal} {\bibinfo
  {journal} {Phys. Rev.}\ }\textbf {\bibinfo {volume} {D97}},\ \bibinfo {pages}
  {103004} (\bibinfo {year} {2018})},\ \Eprint
  {http://arxiv.org/abs/1711.04531} {arXiv:1711.04531 [hep-ph]} \BibitemShut
  {NoStop}%
%%CITATION = ARXIV:1711.04531;%%
\bibitem [{\citenamefont {Cherry}\ and\ \citenamefont
  {Shoemaker}(2019)}]{Cherry:2018rxj}%
  \BibitemOpen
  \bibfield  {author} {\bibinfo {author} {\bibfnamefont {J.~F.}\ \bibnamefont
  {Cherry}}\ and\ \bibinfo {author} {\bibfnamefont {I.~M.}\ \bibnamefont
  {Shoemaker}},\ }\href {\doibase 10.1103/PhysRevD.99.063016} {\bibfield
  {journal} {\bibinfo  {journal} {Phys. Rev.}\ }\textbf {\bibinfo {volume}
  {D99}},\ \bibinfo {pages} {063016} (\bibinfo {year} {2019})},\ \Eprint
  {http://arxiv.org/abs/1802.01611} {arXiv:1802.01611 [hep-ph]} \BibitemShut
  {NoStop}%
%%CITATION = ARXIV:1802.01611;%%
\bibitem [{\citenamefont {Yin}(2019)}]{Yin:2018yjn}%
  \BibitemOpen
  \bibfield  {author} {\bibinfo {author} {\bibfnamefont {W.}~\bibnamefont
  {Yin}},\ }\bibfield  {booktitle} {\emph {\bibinfo {booktitle} {{Proceedings,
  20th International Symposium on Very High Energy Cosmic Ray Interactions
  (ISVHECRI 2018): Nagoya, Japan, May 21-25, 2018}}},\ }\href {\doibase
  10.1051/epjconf/201920804003} {\bibfield  {journal} {\bibinfo  {journal} {EPJ
  Web Conf.}\ }\textbf {\bibinfo {volume} {208}},\ \bibinfo {pages} {04003}
  (\bibinfo {year} {2019})},\ \Eprint {http://arxiv.org/abs/1809.08610}
  {arXiv:1809.08610 [hep-ph]} \BibitemShut {NoStop}%
%%CITATION = ARXIV:1809.08610;%%
\bibitem [{\citenamefont {Fox}\ \emph {et~al.}(2018)\citenamefont {Fox},
  \citenamefont {Sigurdsson}, \citenamefont {Shandera}, \citenamefont
  {M\'esz\'aros}, \citenamefont {Murase}, \citenamefont {Mostaf\'a},\ and\
  \citenamefont {Coutu}}]{Fox:2018syq}%
  \BibitemOpen
  \bibfield  {author} {\bibinfo {author} {\bibfnamefont {D.~B.}\ \bibnamefont
  {Fox}}, \bibinfo {author} {\bibfnamefont {S.}~\bibnamefont {Sigurdsson}},
  \bibinfo {author} {\bibfnamefont {S.}~\bibnamefont {Shandera}}, \bibinfo
  {author} {\bibfnamefont {P.}~\bibnamefont {M\'esz\'aros}}, \bibinfo {author}
  {\bibfnamefont {K.}~\bibnamefont {Murase}}, \bibinfo {author} {\bibfnamefont
  {M.}~\bibnamefont {Mostaf\'a}}, \ and\ \bibinfo {author} {\bibfnamefont
  {S.}~\bibnamefont {Coutu}},\ }\href@noop {} {\  (\bibinfo {year} {2018})},\
  \Eprint {http://arxiv.org/abs/1809.09615} {arXiv:1809.09615 [astro-ph.HE]}
  \BibitemShut {NoStop}%
\bibitem [{\citenamefont {Heurtier}\ \emph {et~al.}(2019)\citenamefont
  {Heurtier}, \citenamefont {Mambrini},\ and\ \citenamefont
  {Pierre}}]{Heurtier:2019git}%
  \BibitemOpen
  \bibfield  {author} {\bibinfo {author} {\bibfnamefont {L.}~\bibnamefont
  {Heurtier}}, \bibinfo {author} {\bibfnamefont {Y.}~\bibnamefont {Mambrini}},
  \ and\ \bibinfo {author} {\bibfnamefont {M.}~\bibnamefont {Pierre}},\ }\href
  {\doibase 10.1103/PhysRevD.99.095014} {\bibfield  {journal} {\bibinfo
  {journal} {Phys. Rev. D}\ }\textbf {\bibinfo {volume} {99}},\ \bibinfo
  {pages} {095014} (\bibinfo {year} {2019})},\ \Eprint
  {http://arxiv.org/abs/1902.04584} {arXiv:1902.04584 [hep-ph]} \BibitemShut
  {NoStop}%
\bibitem [{\citenamefont {Kannike}\ \emph {et~al.}(2020)\citenamefont
  {Kannike}, \citenamefont {Raidal}, \citenamefont {Veermäe}, \citenamefont
  {Strumia},\ and\ \citenamefont {Teresi}}]{Kannike:2020agf}%
  \BibitemOpen
  \bibfield  {author} {\bibinfo {author} {\bibfnamefont {K.}~\bibnamefont
  {Kannike}}, \bibinfo {author} {\bibfnamefont {M.}~\bibnamefont {Raidal}},
  \bibinfo {author} {\bibfnamefont {H.}~\bibnamefont {Veermäe}}, \bibinfo
  {author} {\bibfnamefont {A.}~\bibnamefont {Strumia}}, \ and\ \bibinfo
  {author} {\bibfnamefont {D.}~\bibnamefont {Teresi}},\ }\href@noop {} {\
  (\bibinfo {year} {2020})},\ \Eprint {http://arxiv.org/abs/2006.10735}
  {arXiv:2006.10735 [hep-ph]} \BibitemShut {NoStop}%
\bibitem [{\citenamefont {Fornal}\ \emph {et~al.}(2020)\citenamefont {Fornal},
  \citenamefont {Sandick}, \citenamefont {Shu}, \citenamefont {Su},\ and\
  \citenamefont {Zhao}}]{Fornal:2020npv}%
  \BibitemOpen
  \bibfield  {author} {\bibinfo {author} {\bibfnamefont {B.}~\bibnamefont
  {Fornal}}, \bibinfo {author} {\bibfnamefont {P.}~\bibnamefont {Sandick}},
  \bibinfo {author} {\bibfnamefont {J.}~\bibnamefont {Shu}}, \bibinfo {author}
  {\bibfnamefont {M.}~\bibnamefont {Su}}, \ and\ \bibinfo {author}
  {\bibfnamefont {Y.}~\bibnamefont {Zhao}},\ }\href@noop {} {\  (\bibinfo
  {year} {2020})},\ \Eprint {http://arxiv.org/abs/2006.11264} {arXiv:2006.11264
  [hep-ph]} \BibitemShut {NoStop}%
\bibitem [{\citenamefont {Su}\ \emph {et~al.}(2020)\citenamefont {Su},
  \citenamefont {Wang}, \citenamefont {Wu}, \citenamefont {Yang},\ and\
  \citenamefont {Zhu}}]{Su:2020zny}%
  \BibitemOpen
  \bibfield  {author} {\bibinfo {author} {\bibfnamefont {L.}~\bibnamefont
  {Su}}, \bibinfo {author} {\bibfnamefont {W.}~\bibnamefont {Wang}}, \bibinfo
  {author} {\bibfnamefont {L.}~\bibnamefont {Wu}}, \bibinfo {author}
  {\bibfnamefont {J.~M.}\ \bibnamefont {Yang}}, \ and\ \bibinfo {author}
  {\bibfnamefont {B.}~\bibnamefont {Zhu}},\ }\href {\doibase
  10.1103/PhysRevD.102.115028} {\bibfield  {journal} {\bibinfo  {journal}
  {Phys. Rev. D}\ }\textbf {\bibinfo {volume} {102}},\ \bibinfo {pages}
  {115028} (\bibinfo {year} {2020})},\ \Eprint
  {http://arxiv.org/abs/2006.11837} {arXiv:2006.11837 [hep-ph]} \BibitemShut
  {NoStop}%
\bibitem [{\citenamefont {Bloch}\ \emph {et~al.}(2020)\citenamefont {Bloch},
  \citenamefont {Caputo}, \citenamefont {Essig}, \citenamefont {Redigolo},
  \citenamefont {Sholapurkar},\ and\ \citenamefont {Volansky}}]{Bloch:2020uzh}%
  \BibitemOpen
  \bibfield  {author} {\bibinfo {author} {\bibfnamefont {I.~M.}\ \bibnamefont
  {Bloch}}, \bibinfo {author} {\bibfnamefont {A.}~\bibnamefont {Caputo}},
  \bibinfo {author} {\bibfnamefont {R.}~\bibnamefont {Essig}}, \bibinfo
  {author} {\bibfnamefont {D.}~\bibnamefont {Redigolo}}, \bibinfo {author}
  {\bibfnamefont {M.}~\bibnamefont {Sholapurkar}}, \ and\ \bibinfo {author}
  {\bibfnamefont {T.}~\bibnamefont {Volansky}},\ }\href@noop {} {\  (\bibinfo
  {year} {2020})},\ \Eprint {http://arxiv.org/abs/2006.14521} {arXiv:2006.14521
  [hep-ph]} \BibitemShut {NoStop}%
\bibitem [{\citenamefont {Anchordoqui}\ \emph {et~al.}(2021)\citenamefont
  {Anchordoqui}, \citenamefont {Barger}, \citenamefont {Marfatia},
  \citenamefont {Reno},\ and\ \citenamefont {Weiler}}]{Anchordoqui:2021dls}%
  \BibitemOpen
  \bibfield  {author} {\bibinfo {author} {\bibfnamefont {L.~A.}\ \bibnamefont
  {Anchordoqui}}, \bibinfo {author} {\bibfnamefont {V.}~\bibnamefont {Barger}},
  \bibinfo {author} {\bibfnamefont {D.}~\bibnamefont {Marfatia}}, \bibinfo
  {author} {\bibfnamefont {M.~H.}\ \bibnamefont {Reno}}, \ and\ \bibinfo
  {author} {\bibfnamefont {T.~J.}\ \bibnamefont {Weiler}},\ }\href@noop {} {\
  (\bibinfo {year} {2021})},\ \Eprint {http://arxiv.org/abs/2101.09559}
  {arXiv:2101.09559 [astro-ph.HE]} \BibitemShut {NoStop}%
\bibitem [{\citenamefont {Khlebnikov}\ and\ \citenamefont
  {Tkachev}(1996)}]{Khlebnikov:1996mc}%
  \BibitemOpen
  \bibfield  {author} {\bibinfo {author} {\bibfnamefont {S.}~\bibnamefont
  {Khlebnikov}}\ and\ \bibinfo {author} {\bibfnamefont {I.}~\bibnamefont
  {Tkachev}},\ }\href {\doibase 10.1103/PhysRevLett.77.219} {\bibfield
  {journal} {\bibinfo  {journal} {Phys. Rev. Lett.}\ }\textbf {\bibinfo
  {volume} {77}},\ \bibinfo {pages} {219} (\bibinfo {year} {1996})},\ \Eprint
  {http://arxiv.org/abs/hep-ph/9603378} {arXiv:hep-ph/9603378} \BibitemShut
  {NoStop}%
\bibitem [{\citenamefont {Micha}\ and\ \citenamefont
  {Tkachev}(2003)}]{Micha:2002ey}%
  \BibitemOpen
  \bibfield  {author} {\bibinfo {author} {\bibfnamefont {R.}~\bibnamefont
  {Micha}}\ and\ \bibinfo {author} {\bibfnamefont {I.~I.}\ \bibnamefont
  {Tkachev}},\ }\href {\doibase 10.1103/PhysRevLett.90.121301} {\bibfield
  {journal} {\bibinfo  {journal} {Phys. Rev. Lett.}\ }\textbf {\bibinfo
  {volume} {90}},\ \bibinfo {pages} {121301} (\bibinfo {year} {2003})},\
  \Eprint {http://arxiv.org/abs/hep-ph/0210202} {arXiv:hep-ph/0210202}
  \BibitemShut {NoStop}%
\bibitem [{\citenamefont {Micha}\ and\ \citenamefont
  {Tkachev}(2004)}]{Micha:2004bv}%
  \BibitemOpen
  \bibfield  {author} {\bibinfo {author} {\bibfnamefont {R.}~\bibnamefont
  {Micha}}\ and\ \bibinfo {author} {\bibfnamefont {I.~I.}\ \bibnamefont
  {Tkachev}},\ }\href {\doibase 10.1103/PhysRevD.70.043538} {\bibfield
  {journal} {\bibinfo  {journal} {Phys. Rev. D}\ }\textbf {\bibinfo {volume}
  {70}},\ \bibinfo {pages} {043538} (\bibinfo {year} {2004})},\ \Eprint
  {http://arxiv.org/abs/hep-ph/0403101} {arXiv:hep-ph/0403101} \BibitemShut
  {NoStop}%
\bibitem [{\citenamefont {Lozanov}\ and\ \citenamefont
  {Amin}(2018)}]{Lozanov:2017hjm}%
  \BibitemOpen
  \bibfield  {author} {\bibinfo {author} {\bibfnamefont {K.~D.}\ \bibnamefont
  {Lozanov}}\ and\ \bibinfo {author} {\bibfnamefont {M.~A.}\ \bibnamefont
  {Amin}},\ }\href {\doibase 10.1103/PhysRevD.97.023533} {\bibfield  {journal}
  {\bibinfo  {journal} {Phys. Rev. D}\ }\textbf {\bibinfo {volume} {97}},\
  \bibinfo {pages} {023533} (\bibinfo {year} {2018})},\ \Eprint
  {http://arxiv.org/abs/1710.06851} {arXiv:1710.06851 [astro-ph.CO]}
  \BibitemShut {NoStop}%
\bibitem [{\citenamefont {Gorbunov}\ and\ \citenamefont
  {Shaposhnikov}(2007)}]{Gorbunov:2007ak}%
  \BibitemOpen
  \bibfield  {author} {\bibinfo {author} {\bibfnamefont {D.}~\bibnamefont
  {Gorbunov}}\ and\ \bibinfo {author} {\bibfnamefont {M.}~\bibnamefont
  {Shaposhnikov}},\ }\href {\doibase 10.1007/JHEP11(2013)101,
  10.1088/1126-6708/2007/10/015} {\bibfield  {journal} {\bibinfo  {journal}
  {JHEP}\ }\textbf {\bibinfo {volume} {10}},\ \bibinfo {pages} {015} (\bibinfo
  {year} {2007})},\ \bibinfo {note} {[Erratum: JHEP11,101(2013)]},\ \Eprint
  {http://arxiv.org/abs/0705.1729} {arXiv:0705.1729 [hep-ph]} \BibitemShut
  {NoStop}%
%%CITATION = ARXIV:0705.1729;%%
\bibitem [{\citenamefont {Waxman}\ and\ \citenamefont
  {Bahcall}(1999)}]{Waxman:1998yy}%
  \BibitemOpen
  \bibfield  {author} {\bibinfo {author} {\bibfnamefont {E.}~\bibnamefont
  {Waxman}}\ and\ \bibinfo {author} {\bibfnamefont {J.~N.}\ \bibnamefont
  {Bahcall}},\ }\href {\doibase 10.1103/PhysRevD.59.023002} {\bibfield
  {journal} {\bibinfo  {journal} {Phys. Rev. D}\ }\textbf {\bibinfo {volume}
  {59}},\ \bibinfo {pages} {023002} (\bibinfo {year} {1999})},\ \Eprint
  {http://arxiv.org/abs/hep-ph/9807282} {arXiv:hep-ph/9807282} \BibitemShut
  {NoStop}%
\end{thebibliography}%

\end{document}